\documentclass[12pt,oneside]{article}
\newcommand{\Path }{}
\newcommand{\figs }{}
\usepackage{\Path Paper_style} 
\usepackage{\Path Keywords_WB} 
\usepackage{\Path Environments} 
\usepackage{sidecap}
\usepackage{tikz-cd} 
\usepackage{natbib}
\usepackage{\Path tikzcd2}
\usepackage{subcaption}

\newcommand{\rkhs}{\mathcal{H}} 
\newcommand{\noise}{\mathfrak{y}} 

\begin{document}
	\title{\vspace{-1cm}Image denoising as a conditional expectation}
	\author{Sajal Chakroborty\footnotemark[1], Suddhasattwa Das\footnotemark[2]}
	\footnotetext[1]{Department of Mathematical Sciences, Worcester Polytechnic Institute, Massachusetts, USA}
	\footnotetext[2]{Department of Mathematics and Statistics, Texas Tech University, Texas, USA}
	\date{\today}
	\maketitle
	\begin{abstract} 
		All techniques for denoising involve a notion of a true (noise-free) image, and a hypothesis space. The hypothesis space may reconstruct the image directly as a grayscale valued function, or indirectly by its Fourier or wavelet spectrum. Most common techniques estimate the true image as a projection to some subspace. We propose an interpretation of a noisy image as a collection of samples drawn from a certain probability space. Within this interpretation, projection based approaches are not guaranteed to be unbiased and convergent. We present a data-driven denoising method in which the true image is recovered as a conditional expectation. Although the probability space is unknown apriori, integrals on this space can be estimated by kernel integral operators. The true image is reformulated as the least squares solution to a linear equation in a reproducing kernel Hilbert space (RKHS), and involving various kernel integral operators as linear transforms. Assuming the true image to be a continuous function on a compact planar domain, the technique is shown to be convergent as the number of pixels goes to infinity. We also show that for a picture with finite number of pixels, the convergence result can be used to choose the various parameters for an optimum denoising result.
	\end{abstract}
	
	
	\section{Introduction}\label{sec:intro}
	
	One of the most important areas in the field of image processing is that of image restoration, and the most basic task in image restoration is that of denoising. There has been a gradual shift towards the development of non-parametric techniques \cite{Milanfar2012tour, buades2005review}. Such techniques make minimal assumptions on the image or the noise model. Their aim is to exploit local features on the image, considered both as the graph of a function on the unit square, as well as a noisy manifold in 3-dimensional space.
	
	We present a probabilistic approach to image denoising, which allows a precise formulation of the “true” or noise-free image. This distinguishes this approach to the majority of denoising techniques which assume that the noise-free image is the minimizer of some $L^1$ or $L^2$-based objective function \cite[e.g.]{rudin1992nonlinear, chambolle2004algo, BaiFeng2018image, ng2010solving}, and various variations such as Alternating direction Methods \cite[e.g.]{ng2010solving}, iterative methods \cite[e.g.]{vogel1996iterative}, and multilevel denoising \cite[e.g.]{chan2010multilevel, chan2008fast}. These optimization problems face the usual challenges of local minima, as well as a change in the objective function with different datasets. $L^1$ based approaches may also lead to the numerical artifact of \textit{staircasing}, where blocky distortions appear in regions with smooth gradients \cite{chambolle2004algo, vogel1996iterative}. There is often a need to impose special constraints such as convexity \cite{weiss2009efficient,  ng2010solving}. Our method avoids these issues and artifacts, by taking a probabilistic approach to denoising. It is a non-parametric method which is scalable, convergent and free of assumptions of convexity.
	
	In our probabilistic approach, the noise-free image is the conditional expectation of a random variable, with respect to a natural measure $\mu$. A single noisy image can then be interpreted as a collection of random variable drawn from the distribution given by $\mu$, each sample containing a pixel location along with a noise value. This has the advantage that the objective remains fixed and independent of the noise samples as well as the pixellation of the image. The second advantage is that no foreknowledge is required about this measure $\mu$, nor is it required to have a density function. This is a distinguishing feature of our work from Bayesian approaches to denoising \cite[e.g.]{simoncelli1999bayesian} The conditional expectation is analytically equal to   partial integration with respect to the noise variable. These integrals can be realized conveniently by the application of integral operators. The conditional expectation problem is then converted into a linear regression problem in a reproducing kernel Hilbert space (RKHS). 
	
	\begin{figure} [!t]
		\centering
		\begin{subfigure}[t]{0.48\linewidth}
			\centering
			\includegraphics[width=.98\linewidth]{\figs 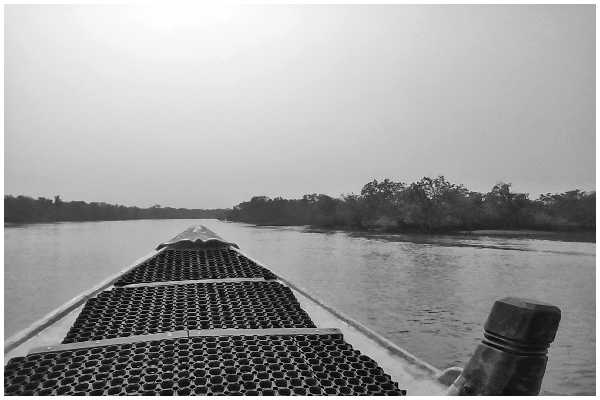}
			\caption{Original}
			\label{fig:shundorbon2:a}
		\end{subfigure}    
		~
		\begin{subfigure}[t]{0.48\linewidth}
			\centering
			\includegraphics[width=.98\linewidth]{\figs 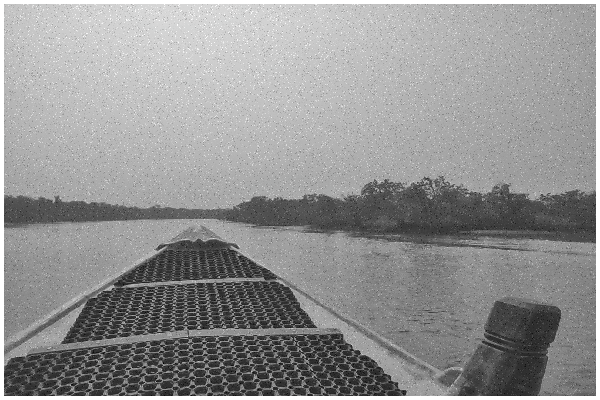}
			\caption{Noisy}
		\end{subfigure}    
		
		\begin{subfigure}[t]{0.48\linewidth}
			\centering
			\includegraphics[width=.98\linewidth]{\figs 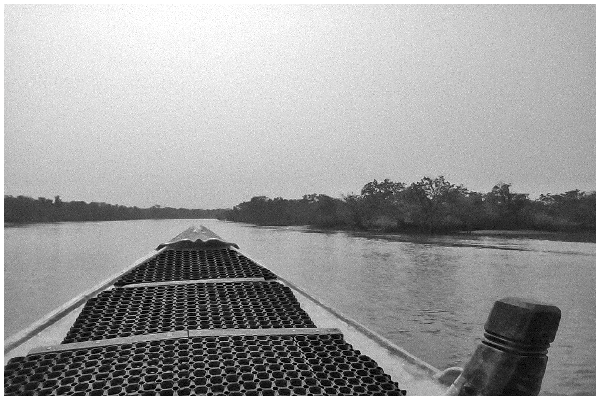}
			\caption{Denoised}
		\end{subfigure}    
		\begin{subfigure}[t]{0.48\linewidth}
			\centering
			\includegraphics[width=.98\linewidth]{\figs 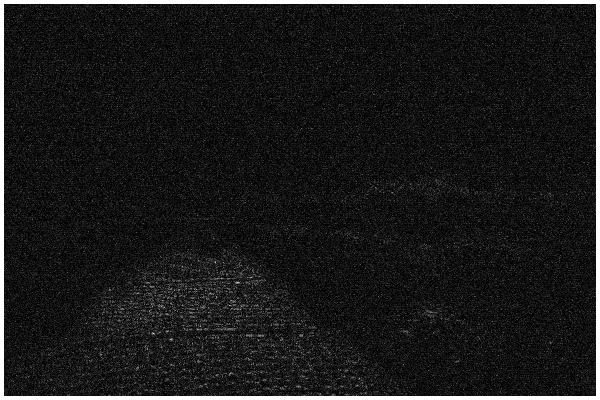}
			\caption{Denoising errors}
		\end{subfigure}    
		\caption{Kernel based denoising - an example. The article presents a data-driven technique for denoising which does not assume any prior distribution for the noise. The denoised image is interpreted as a conditional expectation. The numerical method is thus a means of estimating the conditional expectation from samples distributed uniformly with respect to an underlying measure. The panels show the results of applying this technique to a real world image. The original image has several separate layers, such as a foreground, a background, and other objects in between. Each of the layers have different textures.  The denoised image and its errors from the noise-free image show the effectiveness of this approach to denoising.}
		\label{fig:shundorbon2} 
	\end{figure}
	
	Overall this entire numerical implementation of the procedure is a matrix operation on the image vector, making our technique comparable to a filtering technique, as shown in Figure \ref{fig:linear_filter}. Our use of localized kernels and a RKHS as the hypothesis space lends some of the advantages of local averaging methods. So although our technique is a RKHS-based implementation of a probabilistic principle, it leads to a natural confluence of filtering based approaches \cite{khoury1998closest, milanfar2013symm}, standard kernel regression \cite{takedaEtAl2007kernel}, and local averaging methods \cite[e.g.]{LindenbaumEtAl1994gabor}. We shall explore these connections as the details of our technique unfolds. 
	
	We now state our assumptions precisely. Set $\calX$ to be a compact subset of $[0,1]^2$, the unit square. 
	
	\begin{Assumption} \label{A:1}
		There is a continuous map $\bar{f}: \calX \rightarrow \left[0,1\right]$. 
	\end{Assumption}
	
	Our technique is thus rooted in the assumption of a continuous image. Images are typically piecewise smooth. The techniques based on Assumption \ref{A:1} may be applied separately to the continuous pieces of the image. Such pieces may be extracted by any of the standard techniques of image segmentation \cite[e.g.]{minaee2021image, haralick1985image, ghosh2019und}. This also explains our interpretation of $\calX$ as a subset of the image area.

	Although our understanding of an image is as a continuous map over the domain $\calX$ which is possible a connected set, our denoising algorithm will be based on pixel information, which is based on a grid of pixel locations. To connect the discrete information of the pixellated image with the continuum case we shall interpret the grid to be a sampling of the uniform Lebesgue measure on $\calX$. This measure shall be formally assumed as 
	
	\begin{Assumption} \label{A:3}
		The space $\calX$ is endowed with a probability measure $\mu_X$.
	\end{Assumption}
	
	Assumption \ref{A:3} plays an important role in our convergence analysis. $\mu_X$ shall be required to describe the limiting behavior of our algorithms for infinitely many data points. The function $\bar{f}$ can be interpreted without ambiguity as the true image. Let us set $\calY = \real$, and endow this space with it's Borel sigma-algebra. A variable drawn from $\calY$ will model the noise being added to each pixel. Formally we assume
	
	\begin{Assumption} \label{A:2}
		For each $x\in \calX$ there is a Borel probability measure $\nu_{\calY|x}$ on $\calY$. For $\mu_X$-almost every $x\in \calX$, the measures $\nu_{\calY|x}$ have zero mean.
	\end{Assumption}
	
	\begin{figure} [!t]
		\centering
		\begin{subfigure}[t]{0.48\linewidth}
			\centering
			\includegraphics[width=.98\linewidth]{\figs 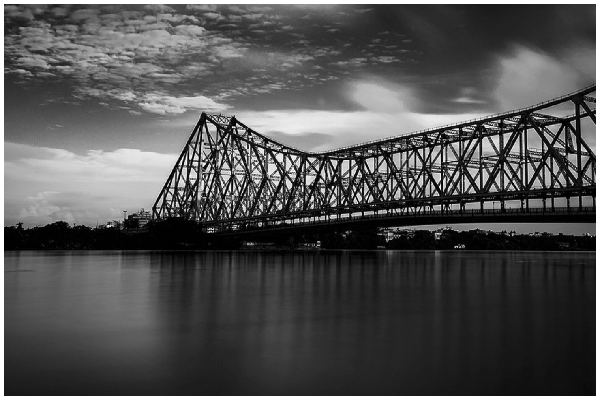}
			\caption{Original}
			\label{fig:howrah_bridge:a} 
		\end{subfigure}    
		~
		\begin{subfigure}[t]{0.48\linewidth}
			\centering
			\includegraphics[width=.98\linewidth]{\figs 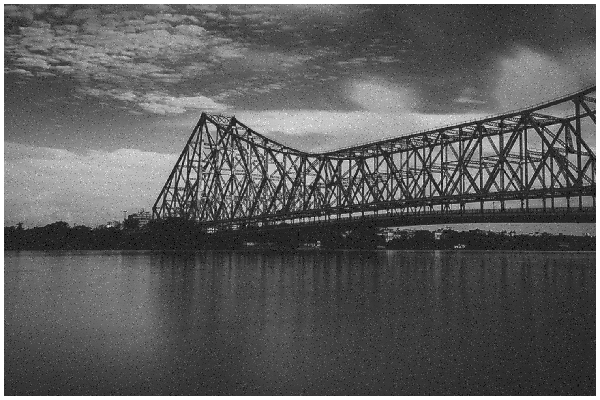}
			\caption{Noisy}
		\end{subfigure}    
		
		\begin{subfigure}[t]{0.48\linewidth}
			\centering
			\includegraphics[width=.98\linewidth]{\figs 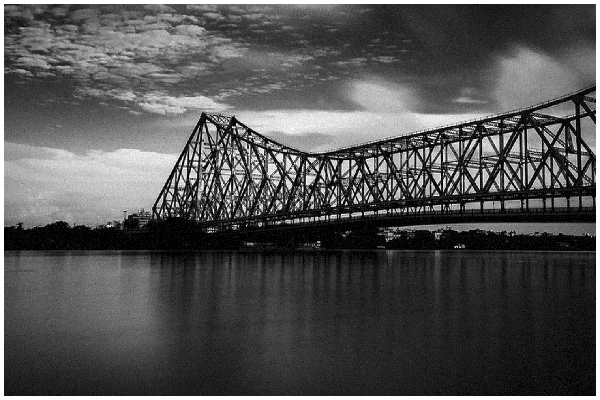}
			\caption{Denoised}
		\end{subfigure}    
		\begin{subfigure}[t]{0.48\linewidth}
			\centering
			\includegraphics[width=.98\linewidth]{\figs 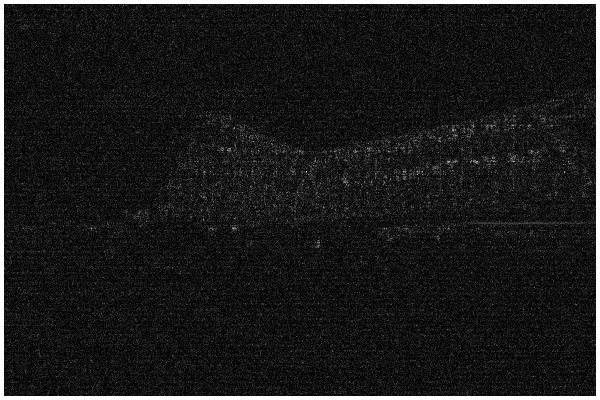}
			\caption{Denoising errors}
		\end{subfigure} 
		\caption{Kernel based denoising - Experiment 4. The goal is the same as in Figure \ref{fig:shundorbon2}. Here the image has a more continuous background and with a non-smooth, low dimensional object in the middle. The RKHS based technique shows the most error at these points of non-smoothness.}
		\label{fig:howrah_bridge} 
	\end{figure}	
	
	Assumptions \ref{A:1} and \ref{A:2} gives us the function
	\begin{equation} \label{eqn:def:img_noise}
		f : \calX \times \calY \to \real, \quad f\left(x, y\right) := \bar{f}\left(x\right)+y
	\end{equation}

	The function $f$ above is a continuous function that also captures the notion of a noisy image, as is the standard practice \cite[e.g.]{Boncelet2009image, rudin1992nonlinear}. There can be several reasons for noisy images, such as mechanical errors, physical damage, and deterioration with time. Based on the source of noise several different noise models have been proposed \cite{Boncelet2009image}, such as gaussian noise, Flicker or Brownian noise, Photon noise and Speckle noise. Noise models such as Gaussian noise, Brown and Flicker noise, assume a noise distribution $\nu_{\calY|x}$, which is independent of the location of $x \in \mathcal{X}$. In that case, each of the $\nu_{\calY|x}$ equals a distribution $\nu$, and $\mu$ is the product $\mu_{X} \times \nu$.
	
	In other words, the measures in Assumptions \ref{A:3} and \ref{A:2} combine into a measure $\mu$ on $\calX \times \calY$, given by:
	\begin{equation} \label{eqn:def:mu}
		\mu := \int_{x\in \calX} \nu_{\calY|x} d\mu_X(x) .
	\end{equation}
	This integration of measures may be interpreted in a weak sense : 
	\[ \int \phi d\mu := \int_{x\in \calX} \left[ \int_{y \in \calY} \phi(x, y) d\nu_{\calY|x}(y) \right] d\mu_X(x) , \quad \forall \phi \in C_c(\calX \times \calY) \]
	Assumptions \ref{A:3} and \ref{A:2} thus imply
	
	\begin{Assumption} \label{A:4}
		There is a probability measure $\mu$ on $\calX \times \calY$ whose projection to $\calX$ is a probability measure $\mu_X$.
	\end{Assumption}
	
	The function $f$ is a random variable on a joint probability space. Now consider any collection $X = \SetDef{x_n}{ 1\leq n \leq N }$ of $N$ points sampled according to $\mu_X$, along with $N$ noise values $\SetDef{ y_n }{ 1\leq n \leq N }$. Each noise level $y_n$ is drawn from $\calY$ according to the conditional distribution $\mu|_{x_n}$. One can consider the pairs
	\begin{equation} \label{eqn:def:noisy_img}
		\paran{ x_n, z_n = f(x_n, y_n) } , \quad 1\leq n \leq N ,
	\end{equation}
	as a noisy image, with grayscale values $z_n$ recorded over locations $x_n$. Conversely any noisy image can be considered to be a sampling of the random variable $f$ according to the distribution $\mu$. This is because 
	\begin{enumerate} [(i)]
		\item the grid of spatial locations at which the pixels are based provide a uniform sampling of $\calX$; and
		\item the additive noise at each pixel are independent. 
	\end{enumerate}
	This interpretation of a noisy image as a sampling of a random variable over a joint probability space $\paran{\calX \times \calY, \mu}$ is the first major principle of our technique. It has been enabled by Assumptions \ref{A:1}, \ref{A:3} and \ref{A:2}, or more simply, by Assumptions \ref{A:1}, and \ref{A:4}.
	
	The second major principle is that the true-image function $\bar{f}$ is the conditional expectation
	\begin{equation} \label{eqn:scheme:1}
		\bar{f}(x) = \mathbb{E}\left[\;f\;|\; X=x \;\right] = \int_{\calY}f\left(x,y\right) d\mu\left(y|x\right). 
	\end{equation}
	The simple relation in equation \eqref{eqn:scheme:1} provides the probabilistic basis of our technique. Conditional expectation is an integral operator which can be approximated well from data \cite{Das2023conditional}. Since every image is a random sampling, it provides an estimate of the conditional expectation $\bar{f}$. Our method of denoising is essentially a method of computing conditional expectation. The principle upheld in \eqref{eqn:scheme:1} is reminiscent of convolution methods for image denoising and deblurring \cite[e.g.]{chan2010multilevel}, but it overcomes the requirement of a fixed pixellation of the continuous image.
	
	A desirable property in any numerical method is consistency. This means that the output of the method converges to a limit, as the information in the data achieves completion. The latter is captured by the notion of \emph{equidistribution}, and is defined later in Section \ref{sec:algo}. The consistency of our numerical method is guaranteed by the consistency of the underlying numerical procedure for conditional expectation.
	
	Any method for a numerical estimation must have a reasonable and apriori notion of the “truth” or target object, which is not based on the data. It is difficult to establish the notion of a true noise-free image in techniques such as total variation minimization \cite{rudin1992nonlinear, chambolle2004algo, weiss2009efficient}, diffusion based \cite[e.g.]{BaiFeng2018image}, and PDE-based denoising \cite{jalab2017image, he2015improved}. In our case the true image exists by assumption. Thus if our technique is applied to different noisy samples of the same noise-free image, one would achieve consistent results. 
	
	\paragraph{Outline} We present the details of the technique in Section \ref{sec:technique}. The algorithmic implementation of this technique is presented next in Section \ref{sec:algo}. We invoke a result from RKHS theory to demonstrate that the results of the algorithm are convergent. We next re-analyze the convergence from a probabilistic viewpoint in Section \ref{sec:cnvrgnc}. We check the efficacy of our methods through some numerical methods in Section \ref{sec:examples}. 
	
	\section{The technique} \label{sec:technique} 
	
	\begin{table}[!t]
		\caption{Basic ingredients of the framework. }
		\begin{tabularx}{\linewidth}{|l|L|}        \hline
			Variable & Interpretation \\ \hline
			$\calX$ & a compact subset of $[0,l] \times [0,w]$ which is the domain of the grayscale image \\ \hline
			$\bar{f}$ & a function $\bar{f} : [0,l] \times [0,w] \to [0,1]$ representing the grayscale image \\ \hline
			$\calY$ & the space $\real$ from which the noise is drawn \\ \hline
			$\mu_{X}$ & uniform probability measure on $\calX$ \\ \hline
			$\mu_{Y}$ & probability measure on $\calY$ representing the noise model \\ \hline
			$N$ & the number of pixels \\ \hline
			$M$ & number of pixels in a sub-grid \\ \hline
			$\delta_2$ & bandwidth of the RKHS kernel \\ \hline
			$\delta_3$ & bandwidth of the local averaging operator \\ \hline
		\end{tabularx}
		\label{tab:param1}
	\end{table}
	
	The presentation and analysis of our method is done most conveniently using the language of kernel integral operators.
	
	\paragraph{Kernel methods} A \emph{kernel} on a space $\calZ$ is a bivariate function $k : \calZ \times \calZ \to \real$, and is interpreted to be a measure of similarity between points on $\calZ$. Bivariate functions such as distance and inner-products are examples of kernels. Kernel based methods offer a non-parametric approach to learning, and have been used with success in many diverse fields such as spectral analysis \cite{DasGiannakis_delay_2019, DasGiannakis_RKHS_2018}, discovery of spatial patterns \cite[e.g.]{GiannakisDas_tracers_2019, DasDimitEnik2020}, and the discovery of periodic and chaotic components of various real world systems \cite[e.g.]{DasMustAgar2023_qpd, DasEtAl2023traffic}, and even abstract operator valued measures \cite{DGJ_compactV_2018}. We shall find use for \emph{gaussian} kernels, defined as
	\begin{equation} \label{eqn:def:GaussK}
		k_{\text{Gauss}, \delta} (x,y) := \exp \paran{ -\frac{1}{\delta} \dist( x, y )^2 } , \quad \forall x,y \in \calZ. 
	\end{equation}
	Gaussian kernels have the important property of being \emph{strictly positive definite}, which means that given any distinct points $x_1, \ldots, x_N$ in $\calZ$, numbers $a_1, \ldots, a_N$ in $\real$, the sum $\sum_{i=1}^{N} \sum_{i=1}^{N} a_i a_j k(x_i, x_j)$ is non-negative, and zero iff all the coefficients $a_i$-s are zero. 
	
	A kernel as simple as a gaussian kernel is rarely used directly. One performs various modifications on the kernel so that it is adapted to the total measure / volume of the space, as well as its local topology. We shall use a markov normalized version of the gaussian kernel \eqref{eqn:def:GaussK}. Given a measure $\beta$ on $\calZ$ and a bandwidth parameter $\delta>0$, we can create a new kernel
	\begin{equation} \label{eqn:def:GaussMrkv}
		k_{ \text{Gauss}, \text{Mrkv} \delta }^{ \beta } : \calZ \times \calZ \to \real , \quad (x, x') \mapsto k_{ \text{Gauss}, \delta } (x, x') / \int_X k_{ \text{Gauss}, \delta } (x, x'') d\beta(x'').
	\end{equation}
	The kernel $k_{ \text{Gauss}, \text{Mrkv} \delta }^{ \beta }$ has the property that for every $x$, the integral with respect to (w.r.t.) the second variable $x'$ is $1$. Thus the kernel $k_{ \text{Gauss}, \text{Mrkv} \delta }^{ \beta }$ mimics a discrete time markov process whose state space is $\calZ$. The bandwidth $\delta$ controls how quickly the value of the kernel decays to zero as $x$ and $x'$ grow apart. This process of modification is an example of an application of an \emph{integral operator}.
	
	\paragraph{Integral operators} Closely associated to kernels are kernel integral operators (k.i.o.). Given a probability measure $\nu$ on $\calZ$, one has an integral operator associated to a continuous kernel $k$, defined as
	\[ K^\nu : L^2(\nu) \to C^r(\calZ) , \quad (K^\nu \phi)(x) := \int_{\calX} k(x,y) \phi(y) d\nu(y) . \]
	If the kernel $k$ is $C^r$, then its image set will also be $C^r$ functions. For this reason, k.i.o.-s are also known as smoothing operators. In fact, under mild assumptions, k.i.o.-s embed functions in $L^2(\nu)$ into function spaces of higher regularity, called \emph{RKHS}. Recall that a kernel $k$ is \emph{symmetric} if for every $x,x'\in X$, $k(x,x') = k(x',x)$. Symmetric kernels allow the use of tools from RKHS theory, which we review shortly.
	
	In this short review of kernel integral operators we have assumed $\calZ$ to be a general topological space. Henceforth, the space $\calX$ from Assumption \ref{A:1} plays the role of $\calZ$. 
	
	\begin{figure} [!t]
		\centering
		\begin{subfigure}[t]{0.48\linewidth}
			\centering
			\includegraphics[width=.98\linewidth]{\figs 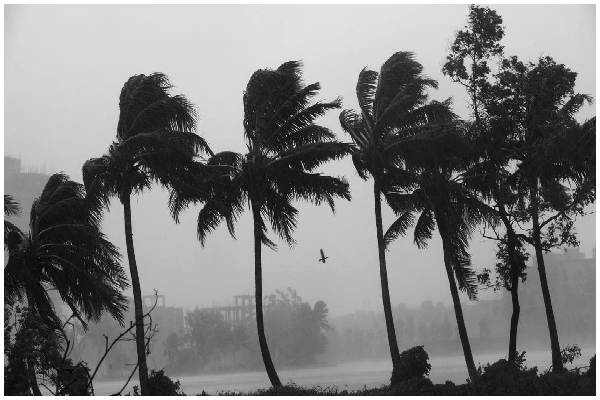}
			\caption{Original}
			\label{fig:jhor:a} 
		\end{subfigure}    
		\begin{subfigure}[t]{0.48\linewidth}
			\centering
			\includegraphics[width=.98\linewidth]{\figs 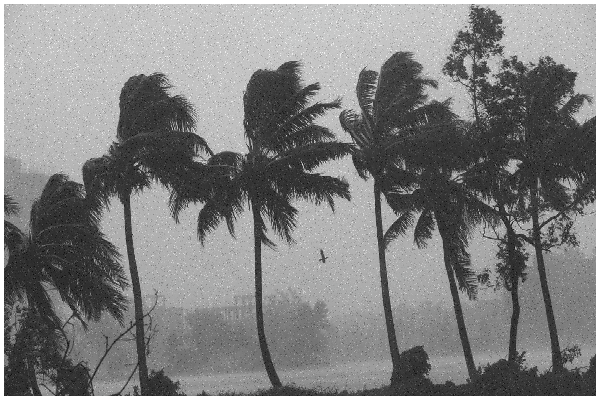}
			\caption{Noisy}
		\end{subfigure}    
		
		\begin{subfigure}[t]{0.48\linewidth}
			\centering
			\includegraphics[width=.98\linewidth]{\figs 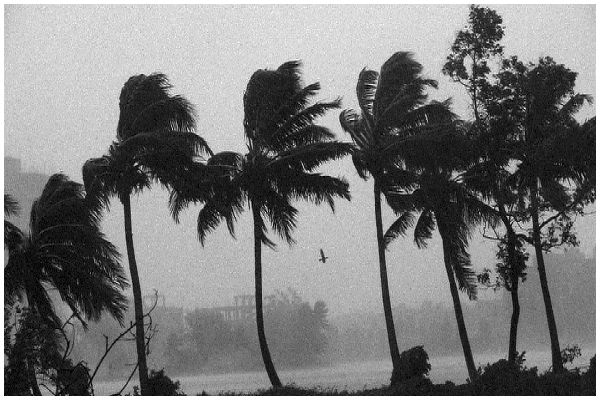}
			\caption{Denoised}
		\end{subfigure}    
		\begin{subfigure}[t]{0.48\linewidth}
			\centering
			\includegraphics[width=.98\linewidth]{\figs 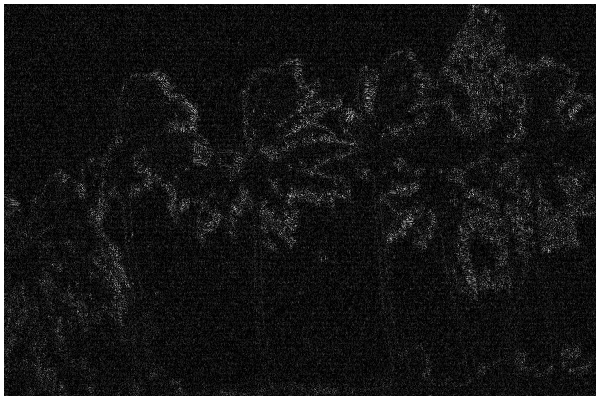}
			\caption{Denoising errors}
		\end{subfigure} 
		\caption{Kernel based denoising - Experiment 5. The goal is the same as in Figures \ref{fig:shundorbon2} and \ref{fig:howrah_bridge}. Here the image is more smooth, but has several objects with fractal outlines. These outlines are reproduced in the heatmap plot of the error. }
		\label{fig:jhor} 
	\end{figure}	
	
\begin{figure}[!t]
		\centering
		\begin{tikzpicture}[scale=0.7, transform shape]
			\node [style={rect3}] (n1) at (1.0\columnA, 0.5\rowA) { Kernel integral operator applied over joint space and noise domain : $\tilde{ \mathbb{P} }^{\mu} f$ };
			\node [style={rect3}] (n2) at (2.5\columnA, 0.5\rowA) { Kernel integral operator over space domain applied to true image :  $\mathbb{P}^{\mu_X} \bar{f}$ };
			\node [style={rect3}] (n3) at (2.5\columnA, 2.2\rowA) { $L^2(\mu_X)$ pre-image \[ \paran{ \mathbb{P}^{\mu} \mathbb{K}^{\mu_X} + \theta }^{-1} \mathbb{P}^{\mu_X} \bar{f} \] };
			\node [style={rect3}] (n4) at (1\columnA, 2.2\rowA) { RKHS approximation of denoised image \[ \begin{split} &\mathbb{K}^{\mu_X} \paran{ \mathbb{P}^{\mu} \mathbb{K}^{\mu_X}  + \theta }^{-1} \cdot \\ & \quad \cdot \mathbb{P}^{\mu_X} \bar{f} \end{split} \] };
			\node [style={rect2}] (S0) at (-0.2\columnA, 1.5\rowA) { \textbf{Noisy image} : a collection \[\SetDef{ (x_n, z_n) }{1 \leq n \leq N} \] of pixel locations $x_n$ and noisy pixel values $z_n$ };
			\node [style={rect2}] (S1) at (-0.2\columnA, -0.8\rowA) { \textbf{Step 1.} Compute $N\times N$ markov kernel matrix \[ \mathbb{P}_{i,n} := p \paran{ x_i, x_n } \] for each $1\leq i, n \leq N$ };
			\node [style={rect2}] (S2) at (1\columnA, -1.9\rowA) { \textbf{Step 2.} Combining all the pixel information using kernel matrix \[ \hat{z}_i := \sum_{n=1}^{N} \mathbb{P}_{i,n} z_n \] for each $1\leq i \leq N$ };
			\node [style={rect2}] (S3) at (2.1\columnA, -1.9\rowA) { \textbf{Step 3.} Compute $N\times M$ kernel matrix \[ \mathbb{K}_{i, j} := k \paran{ x_i, x_j } \] for each $1\leq i \leq N$ and $1\leq j \leq M$};
			\node [style={rect2}] (S4) at (3.2\columnA, -1.9\rowA) { \textbf{Step 4.} Compute the $\theta$-regularized least squares solution to \[ \mathbb{P} \mathbb{K} a = \hat{z} \] };
			\node [style={rect4}] (S4b) at (3.2\columnA, 3.9\rowA) { $a_{N, M, \theta} \in \real^{M}$ }; 
			\node [style={rect5}] (S5) at (1\columnA, 3.9\rowA) { \textbf{Step 5.} Reconstruct the $n$-th pixel as $\sum_{j=1}^{M} \mathbb{K}_{n, j} \paran{a_{N, M, \theta}}_j$ };
			\draw[-to] (n1) to (n2);
			\path [-to] (n1) -- node [midway, below , align=center ] {Equation \eqref{eqn:scheme:2}} (n2);
			\path [-to] (n1) -- node [midway, above , align=center ] {=} (n2);
			\path [-to] (S2) -- node [midway, right , align=center ] {$N\to \infty$} (n1);
			\draw[-to] (S2) to (n1);
			\draw[-to] (S0) to (S1);
			\draw[-to] (S1) to (S2);
			\draw[-to] (S2) to (S3);
			\draw[-to] (S3) to (S4);
			\draw[-to] (S4) to (S4b);
			\draw[-to] (S4b) to (n3);
			\path [-to] (S4b) -- node [midway, left , align=center ] {$M\to \infty$} (n3);
			\draw[-to] (n2) to (n3);
			\path [-to] (n2) -- node [midway, left , align=center ] {invert} (n3);
			\draw[-to] (n3) to (n4);
			\path [-to] (n3) -- node [midway, below , align=center ] {smoothen} (n4);
			\draw[-to] (S4b) to (S5);
		\end{tikzpicture}
		\caption{Mathematical principle for denoising. The algorithmic steps (white) as well as the theoretical rationale (grey) are outlined in this diagram. Our approach is based on an interpretation of the joint color and pixel location as a sampling of a probability space. That way we convert the denoising problem into the task of finding a conditional expectation. In the above diagram, $\bar{f}$ and $f$ are functional representations of the true and noisy images respectively. See \eqref{eqn:def:img_noise} and \eqref{eqn:scheme:1} for more details. The kernel based approach for estimating $\bar{f}$ from pixel information is presented in Section \ref{sec:technique}. The steps outlined here pertain to the last algorithmic step in the flowchart of Figure \ref{fig:outline:2}. }
		\label{fig:outline:1}
	\end{figure}
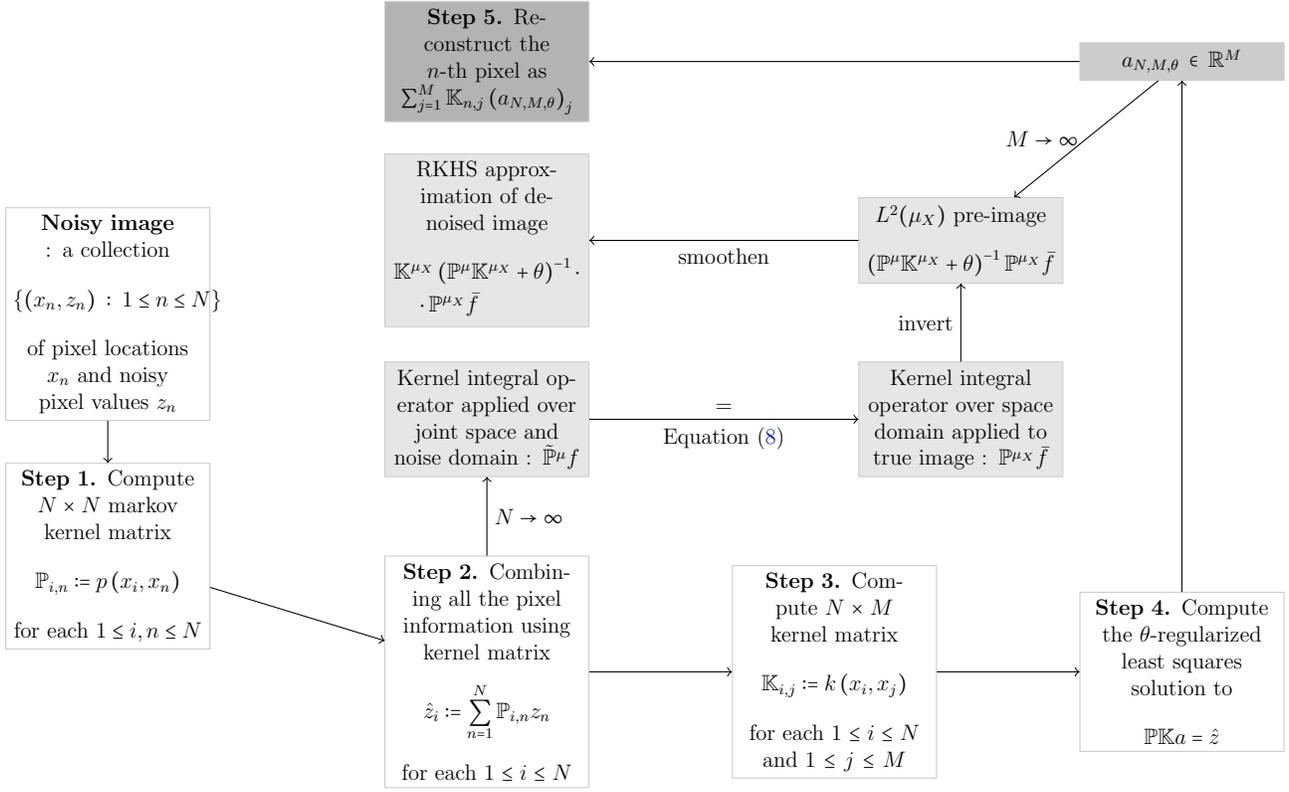

	\paragraph{Denoising} Let $\phi : \calX \to \real$ be any continuous function. This can be extended to a function 
	\[ \tilde{\phi}(x,y) := \phi(x), \]
	on $\calX \times \calY$. Now note that
	\begin{equation} \label{eqn:odo89}
		\begin{split}
			\int_{\calX \times \calY} \tilde\phi(x,y) f(x,y) d\mu(y) &= \int_{x\in \calX} \left[ \int_{y\in \calY} \tilde\phi(x,y) f(x,y) d\mu(y|x)  \right] d\mu_X(x) \\
			&= \int_{x\in \calX} \phi(x) \left[ \int_{y\in \calY} f(x,y) d\mu(y|x)  \right] d\mu_X(x)  \\
			&= \int_{x\in \calX} \phi(x) \bar{f}(x) d\mu_X(x) . 
		\end{split}
	\end{equation}
	Equation \eqref{eqn:odo89} relates the integral of $f$ with $\tilde{\phi}$, to the integral of its conditional expectation $\bar{f}$ with $\phi$. Now let $p : \calX \times \calX \to \real$ be any markov transition function, and $\tilde{p} : \calX \times \calX \times \calY \to \real$ be the trivial extension
	\[ \tilde{p}( x, x', y) := p(x, x') . \]
	Let us fix a point $x_0 \in \calX$. Then the kernel sections $p(x_0, \cdot)$ and $\tilde{p}( x_0, \cdot, \cdot )$ are continuous functions on $\calX$ and $\calX \times \calY$ respectively. Then replacing $\phi$ and $\tilde\phi$ in \eqref{eqn:odo89} by $p(x_0, \cdot)$ and $\tilde{p}( x_0, \cdot, \cdot )$ gives
	\[ \int_{\calX \times \calY} \tilde{p}( x_0, x, y ) f(x,y) d\mu(x,y) = \int_{x\in \calX} p(x_0, x) \bar{f}(x) d\mu_X(x) . \]
	This equality can be concisely restated in terms of kernel integral operators as 
	\begin{equation} \label{eqn:scheme:2}
		\boxed{ \tilde{ \mathbb{P} }^{\mu} f = \mathbb{P}^{\mu_X} \bar{f} }.
	\end{equation}
	The identity in \eqref{eqn:scheme:2} represents the mathematical principle underlying our numerical method. The left hand side (LHS) involves a k.i.o. $\tilde{ \mathbb{P} }^{\mu}$ on the space $\calX \times \calX \times \calY$ applied to the random variable $f$. The RHS involves a k.i.o. $\mathbb{P}^{\mu_X}$ on the space $\calX \times \calX$ applied to the conditional expectation $\bar{f}$. Figure \ref{fig:outline:1} presents how \eqref{eqn:scheme:2} lies at the heart of our technique. The LHS is approximated entirely from data. Equation \eqref{eqn:scheme:2} sets this equal to the true image $\bar{f}$ transformed by a smoothing operator $\mathbb{P}^{\mu_X}$. Thus \eqref{eqn:scheme:2} serves as the bridge between data and the underlying noiseless image. 
	
	The rest of the procedure is aimed at inverting this operator theoretic transformation. Thus the task is now that of linear regression. The expectation $\bar{f}$ is assumed to be in $C(\calX)$. Therefore, $C(\calX)$ is a logical choice for a search space. However the space $C(\calX)$ carries some technical problems as a hypothesis space. It lacks a Hilbert space structure and thus the notion of orthogonal projections. The space $C(\calX)$ can be embedded into $L^2(\mu_X)$. The latter is a Hilbert space, but its members are not true functions but equivalence classes of functions. For this reason, we adopt an \emph{RKHS} as our search space.
	
	\paragraph{RKHS} A reproducing kernel Hilbert space or RKHS is a Hilbert space of continuous functions, in which pointwise evaluations are bounded linear functionals. Any continuous symmetric, strictly positive definite kernel $k$ (such as \eqref{eqn:def:kdiff}) induces an RKHS which contains linear sums of the form $\sum_{n=1}^N a_n k(\cdot, x_n) $, in which the inner product  of any two finite sums is given by
	\[ \bracketBig{ \sum_{n=1}^N a_n k(\cdot, x_n) , \sum_{m=1}^M b_m k(\cdot, y_m) }  = \sum_{n=1}^{N} \sum_{m=1}^{M} a_n^* b_m k(x_n, y_m).  \]
	Here $a^*_m$ denotes the complex conjugate of the coefficient $a_m$.
	This notion of a \emph{pre-inner product} can be given completion in a Hilbert space, which turns out to be the RKHS. Readers can find full details in common sources such as \cite[e.g.]{Paulsen2016}.  The functions $k(\cdot, x_n)$ are called the \emph{sections} of the kernel $k$. Although it is difficult to define the exact span of the RKHS, its use is made simple by the fact that the kernel sections are members of the RKHS and span the RKHS. One of the defining properties of RKHS is the \emph{reproducing} property :
	\[ \bracketBig{ k(\cdot, x) , f } = f(x) , \quad \forall x\in X, \, \forall f\in \rkhs . \]
	Pointwise evaluation is a desirable property in data-driven techniques which is lost when the hypothesis space is an $L^p$ or Sobolev space. Another useful tool is the availability of inner products and therefore a notion of orthogonality, which is absent in non-Hilbert hypothesis spaces such as spaces of $C^r$ functions. RKHS combines both these features. Moreover, under mild conditions \cite{SriperumbudurEtAl2010, MicchelliEtAl2006} $\rkhs$ is an \emph{universal} approximator, i.e., it is dense in $C(\calX)$. We shall therefore assume that
	
	\begin{Assumption} \label{A:5}
		There is continuous, strictly positive definite kernel $k$ on $\calX$.
	\end{Assumption}
	
	When an RKHS is used as the hypothesis space in a learning problem, the target function is assumed to be a finite sum $\sum_{n=1}^N a_n k(\cdot, x_n) $ of the kernel sections. Let $\nu$ be any probability measure on $X$, and $K^\nu$ be the kernel integration operator associated to $k$ and $\nu$. Then it is well known that the image of $K^\nu$ lies in $\rkhs$. We denote this image as $\rkhs_\nu$.  Finite sums of kernel sections can also be interpreted as images of kernel integral operators, using the notion of \emph{sampling measures}.
	
	\paragraph{Sampling measures} Given a measure $\alpha$ on a space $\calZ$, a sequence of points $\braces{z_n}_{n=1}^{\infty}$ is said to be \emph{equidistributed} w.r.t. $\alpha$ if the sequence $\frac{1}{N} \sum_{n=1}^{N} \delta_{z_n} $ of empirical measures converges weakly to $\alpha$. Suppose $\braces{x_n}_{n=1}^{\infty}$ is a sequence of points on $\calX$ equidistributed w.r.t. $\mu_X$. Similarly let $\braces{y_n}_{n=1}^{\infty}$ be a sequence of noise samples equidistributed w.r.t. $\mu_Y$. Let us fix a size $N\in \num$. Let $\beta_X$ and $\beta_Y$ denote the empirical measures constructed from $\braces{x_n}_{n=1}^{N}$ and $\braces{y_n}_{n=1}^{N}$ respectively. Similar to \eqref{eqn:def:mu} one can construct a measure $\beta$ on $\calX \times \calY$, which in this case is 
	\[ \beta := \frac{1}{N} \sum_{n=1}^{N} \delta_{ (x_n, y_n) } .\]
	Note that as $N\to \infty$ the measure $\beta$ converges weakly to $\mu$. Let $\nu$ be a discrete measure $\nu = \sum_{n=1}^{N} w_n \delta_{x_n}$, i.e., an aggregate of Dirac-delta measures supported on discrete points $x_n$ along with weights $w_n \geq 0$ which sum to $1$. Then $\rkhs_\nu$ is precisely the span of the kernel sections $\SetDef{ k(\cdot, x_n) }{ n=1,\ldots,N }$. The Hilbert space $L^2(\nu)$ then becomes isomorphic to a finite dimensional Hilbert space. Two functions on $X$ will be equal in $L^2(\nu)$ iff they are equal on the sample points.
	
	\paragraph{Denoising as regression} The denoising problem can be adjusted in the following manner. We fix a probability measure $\nu$ on $X$. The goal now is to find an $L^2(\nu)$ function $a$ such that $\bar{f} \approx K^\nu a$. Combined with \eqref{eqn:scheme:2}, our objective becomes 
	\begin{equation} \label{eqn:scheme:3}
		\mbox{Find } a\in L^2(\nu) \quad \mbox{ s.t. } \quad \tilde{ \mathbb{P} }^{\mu} f \approx \mathbb{P}^{\mu_X} K^\nu a .
	\end{equation}
	The operators $\tilde{ \mathbb{P} }^{\mu}$ and $\mathbb{P}^{\mu_X}$ are based on the measure $\mu$ and are thus of infinite rank. In a data-driven approach each of the measures $\mu$, $\mu_X$ and $\nu$ must be replaced by sampling measures. 
	
	\begin{figure}[!t]
		\centering
		\includegraphics[width=0.95\linewidth, height=0.5\textheight, keepaspectratio]{\figs 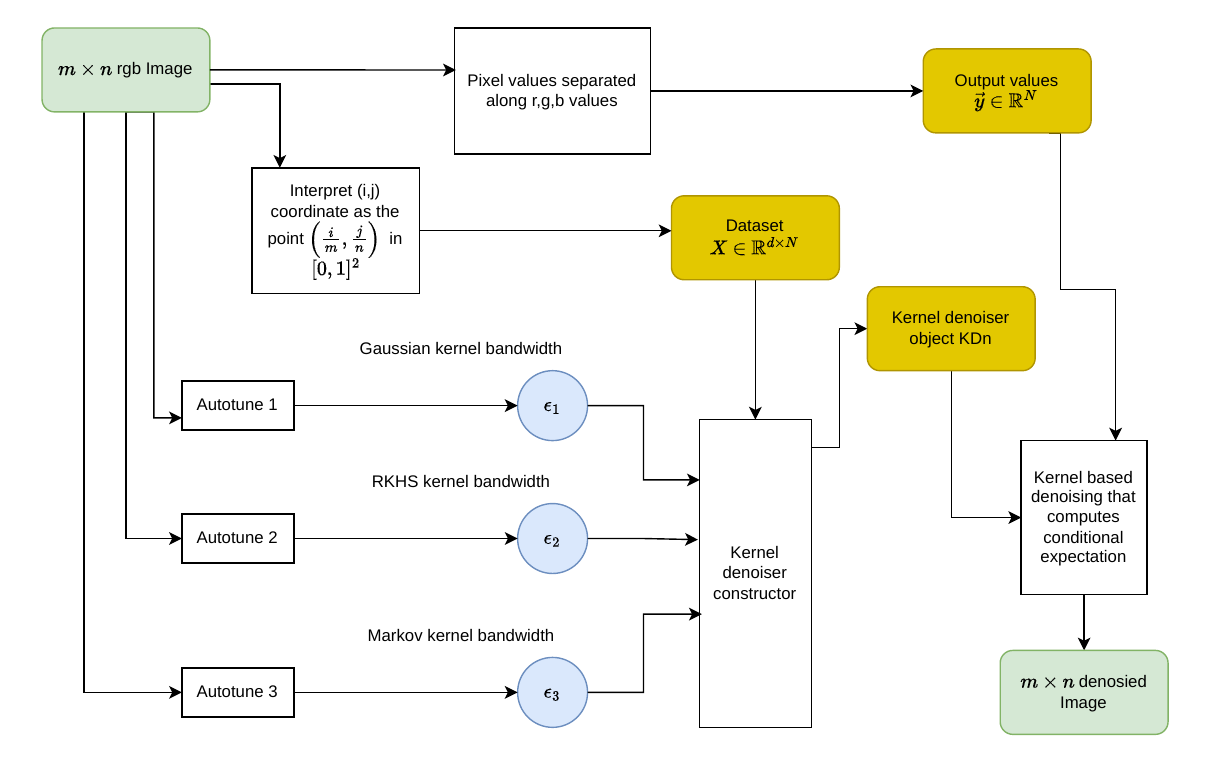}
		\caption{Outline of the algorithms. All the algorithmic steps involved in the denoising algorithm are presented, along with the intermediate objects. The input and output objects are shown in green, algorithmic parameters in blue, and intermediate objects as yellow. The steps outlined in Figure \ref{fig:outline:1} pertain to the last algorithmic step in this flowchart. It is also described in detail in Algorithm \ref{algo:1}. The auto-tuning process is described in Algorithm \ref{algo:2}. The flowchart reveals that the kernel denoiser labelled as KDn is based only on the grid size, and is independent of the color information. KDn combines linearly with the color information in the final step. The details of that step is also depicted separately in Figure \ref{fig:linear_filter}.}
		\label{fig:outline:2}
	\end{figure}
	
	Although the equation \eqref{eqn:scheme:3} provides full statement of the denoising problem, the measures involved, namely $\mu$ and $\mu_X$ are usually inaccessible to data. The measure $\mu$ will be approximated by a suitable chosen measure $\beta$, the projection $\mu_x$ separately by two measures $\beta_X$ and $\nu$. Using these measures we reformulate \eqref{eqn:scheme:3} as a least squares problem
	\[ \tilde{ \mathbb{P} }^{\beta} f \approx \mathbb{P}^{\beta_X} K^{\nu} a .  \]
	Let $\theta>0$ be a fixed constant. Given a generic linear regression problem $Ax = y$, the $\theta$-regularized least squares solution will be the vector 
	\begin{equation} \label{eqn:def:ridge}
		\SqBrack{ A^T A + \theta }^{-1} A^T y .
	\end{equation}
	Using the regularization procedure \eqref{eqn:def:ridge}, the regression \eqref{eqn:scheme:3} becomes
	\begin{equation} \label{eqn:scheme:4}
		\bar{f} \approx \hat{f}_{\beta, \nu, \theta}, \quad \hat{f}_{\beta, \nu, \theta} := K^\nu a_{\beta, \nu, \theta} , \quad a_{\beta, \nu, \theta} := \begin{array}{c} {} \\ \arg\min \\ \theta-\text{reg} \end{array} 
		\SetDef{ \norm{ \mathbb{P}^{\beta_X} K^{\nu} a - \tilde{ \mathbb{P} }^{\beta} f }_{L^2(\nu)} }{ a \in L^2(\nu) } .
	\end{equation}
	Thus the exact relation in \eqref{eqn:scheme:2} has undergone substitutions to yield the approximate relations \eqref{eqn:scheme:3} and \eqref{eqn:scheme:4}. The following result states that the approximate estimate is still convergent :
	
	\begin{lemma} \label{lem:scheme:6}
		\cite[Thm 1]{Das2023conditional} Suppose Assumptions \ref{A:1}, \ref{A:4} and \ref{A:5} hold. Let $\hat{f}_{\beta, \nu, \theta}$ be the estimate according to \eqref{eqn:scheme:4}, where $\beta$ is a measure on $\calX\times \calY$, $\beta_X$ is its projection into $\calX$, and $\nu$ is a probability measure on $\calY$. Suppose further that the conditional expectation $\bar{f}$ lies in $\rkhs$. Then
		\begin{equation} \label{eqn:scheme:6}
			\lim_{\nu \to \mu_X} \lim_{\beta \to \mu, \theta\to 0^+} \norm{ \hat{f}_{\beta, \nu, \theta} - \bar{f} }_{C(\calX)} = 0.
		\end{equation}
	\end{lemma}
	
	Lemma \ref{lem:scheme:6} provides the theoretical justification of our method. Note that $a_{\beta, \nu, \theta}$ from \eqref{eqn:scheme:4} is a function in $L^2(\nu)$. If $\nu$ is a sampling measure, then $a_{\beta, \nu, \theta}$ is a finite vector. Irrespective of the nature of $L^2(\nu)$, the kernel integral operator $K^\nu$ converts it into a smooth function on $\calX$. In general for every vector $a\in L^2(\nu)$, 
	\begin{equation} \label{eqn:def:os}
		\paran{ K^\nu a }(x) = \int_{\calX} k(x, x') a(x') d\nu(x') ,\quad \forall x\in \calX.
	\end{equation}
	We next present the data-driven procedure for solving \eqref{eqn:scheme:6}. An outline is provided in Figure \ref{fig:outline:2}. 
	
	\section{Data-driven implementation} \label{sec:algo} 
	
	All the algorithms are based on the assumption that the image is a collection of $N$ pairs $\SetDef{ \paran{ x_n, z_n } }{ 1 \leq n \leq N }$ as described in \eqref{eqn:def:noisy_img}. We then choose the following sequence of measures : 
	
	\begin{enumerate} [(i)]
		\item We choose $\beta$ to be the empirical sampling measure $\beta := \frac{1}{N} \sum_{n=1}^{N} \delta_{(x_n, y_n)}$ on $\calX \times \calY$.
		\item Its projection on $\calX$ is the empirical sampling measure $\beta_X := \frac{1}{N} \sum_{n=1}^{N} \delta_{(x_n, y_n)}$.
		\item Fix an integer $M<N$, we call this the sub-sampling parameter. We set $\nu$ to be the empirical sampling measure $\frac{1}{M} \sum_{m=1}^{M} \delta_{x'_m}$ on $\calX$. Here $x'_1, \ldots, x'_M$ is a collection of $M$ equally spaced points among $x_1, \ldots, x_M$.
	\end{enumerate}
	
	Due to our choice of $\beta_, \beta_X$ and $\nu$, the spaces $L^2(\nu), L^2(\beta), L^2(\beta_X)$ are finite dimensional, and the operators $\mathbb{P}^{\beta_X}$ and $K^{\nu}$ can be represented by finite matrices. Thus the optimization \eqref{eqn:scheme:4} is in a finite dimensional vector space. It has an explicit solution using the pseudo-inverse. The following algorithm computes the solution to \eqref{eqn:scheme:4} : 
	
	\begin{algo} \label{algo:1}
		RKHS representation of conditional expectation.
		\begin{itemize}
			\item \textbf{Input.} A sequence of pairs $\SetDef{(x_n, z_n)}{ n=1,\ldots,N}$ with $x_n\in\real^d$ and $z_n\in \real$.
			\item \textbf{Parameters.}
			\begin{enumerate} [(i)] 
				\item Choice of RKHS kernel $k:\real^d \times \real^d \to \real^+$.
				\item Choice of a Markov kernel $p:\real^d \times \real^d \to \real^+$.
				\item Smoothing parameter $\theta>0$.
				\item Sub-sampling parameter $M\in\num$ with $M<N$.
				\item Regularization parameter $\theta$.
			\end{enumerate}
			\item \textbf{Output.} A vector $\vec{a} = \paran{a_1, \ldots , a_M} \in \real^M$ such that
			\[ (E^{\alpha} f)(x) \approx \sum_{m=1}^{M} a_m k(x, x_m) , \quad \forall x \in \real^d . \]
			\item \textbf{Steps.}
			\begin{enumerate}
				\item Compute a gaussian markov kernel matrix using \eqref{eqn:def:GaussMrkv} :
				\begin{equation} \label{eqn:GM}
					\Matrix{G_{\delta}} \in \real^{N\times N}, \quad \Matrix{G_\delta}_{i,j} = k_{ \text{Gauss}, \delta }^{ \text{symm}, \beta } (x_i, x_j) .
				\end{equation}
				\item Compute the markov matrix $\Matrix{P^{\beta_X}} \in \real^{N\times N}$ as $\Matrix{P^{\beta_X}}_{i,j} := p(x_i, x_j)$
				\item Compute the kernel matrix $\Matrix{K^{\nu}}\in \real^{N\times M}$ as  $\Matrix{K^{\nu}}_{i,j} = k(x_i, x_j)$.
				\item Find a vector $\vec{a} \in \real^M$ as the $\theta$-regularized least-squares solution to the equation
				\[\Matrix{P^{\beta_X}} \Matrix{K^{\nu}} \vec{a} = \Matrix{P^{\beta_X}} \Matrix{G_\delta} \vec{z} . \]
			\end{enumerate}
		\end{itemize}
	\end{algo}
	
	Note that Algorithm \ref{algo:1} is a general algorithm that performs denoising by computing the conditional expectation. In our case, the $x_n$ in Algorithm \ref{algo:1} represent pixel coordinates of the image, and the $z_n$ represent the noise grayscale values on these pixels. The outline of the technique is presented in Figures \ref{fig:outline:1} and \ref{fig:outline:2}. The various parameters and objects are summarized in Table \ref{tab:param1}.  
	
	\paragraph{Denoising in patches} An image is a data-point in a high-dimensional Euclidean space $\real^{mn}$. Thus kernel based image processing methods may require significant computational resources, which can become a stumbling block. We have proposed computing \(\hat{f}_{\beta, \nu, \theta}\) in \eqref{eqn:scheme:4} following the steps in Algorithm \ref{algo:1}. We later provide an explicit decomposition of the overall procedure as a sequence of matrix compositions, in \eqref{eqn:def:err:3}. For computational feasability, all the matrices involved should preferably be sparse. We promote sparsity by a suitable choice of the bandwidths $\delta_2$ and $\delta_3$. The parameter \(\delta_3\) in equation \eqref{eqn:def:err:3} represents bandwidth of the  Markov averaging operator, which determines the number of neighboring pixels to perform the averaging operation. A larger value of \(\delta_3\) reduces local variation, consequently makes the convergence faster. However, a larger value of \(\delta_3\) will reduce the sparsity and make the computational procedure very slow. Therefore, we consider a reasonably large value for \(\delta_3\). We explain its selection procedure in Algorithm \ref{algo:2}. 
	
	Since $\delta_3$ cannot be reduced to arbitrary values in practical situations, there is a need for efficient memory management by reducing the size of the matrices. We propose to achieve this by partitioning the image into smaller chunks denoising one chunk of the image at a time. More specifically, our method employs a subcover for each image block, denoises it and then moves to the next block. Each subcover is a bounded set. Note that these subcovers overlap, which ensures that essential information in the image such as sharp boundaries or edges is preserved during denoising.
	
	We next explain this procedure in more detail. Let $\calU$ be any bounded set, $h>0$ and $\delta_1>0$ be any positive constants. Then one has a function $\phi : \real^2 \to [0,1]$ defined as
	\begin{equation} \label{eqn:dpd9k}
		\phi(x) = \phi(x; \calU, h) := \begin{cases}
			0 & \mbox{ if } x\notin \overline{\calU} \\
			1 - \exp\paran{ - \dist(x, \partial \calU)^2 / h^2 \delta_1 } & \mbox{ if } x\in \overline{\calU}
		\end{cases}
	\end{equation}
	Next let $\calU_1, \ldots, \calU_m$ be a covering of the space $\calX$ by bounded sets. Let the diameters of the pieces of the cover be $h_i := \diam( \calU_i )$. One can use Algorithm \ref{algo:1} to approximate the conditional expectations of $f$ restricted to the spaces $\calU_i$. Call these estimates $\hat{f}_i$. These estimates may be aggregated together in the following manner.
	\begin{equation} \label{eqn:aed54g}
		\mathfrak{s}(x) := \sum_{i=1}^{m} \phi \paran{x; \calU_i, h_i} , \quad \hat{f} (x) = \frac{1}{ \mathfrak{s}(x) } \sum_{i=1}^{m} \hat{f}_i (x) \phi \paran{x; \calU_i, h_i} .
	\end{equation}
	The schemes in \eqref{eqn:dpd9k} and \eqref{eqn:aed54g} are implemented as follows :
	
	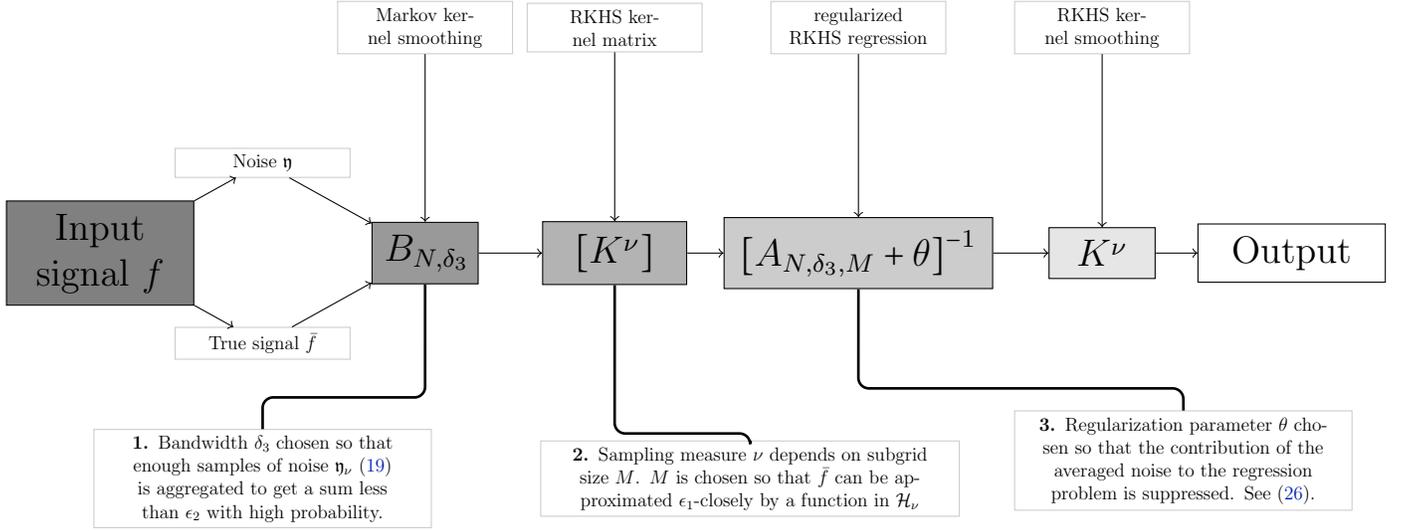
\begin{figure}[!t]
		\centering
		\begin{tikzpicture}[scale=0.6, transform shape]
			\node [text width=0.3\columnA, text centered, minimum height=0.2\rowA, shape=rectangle, draw=black, fill=ChhaiE, scale=2.0] (n7) at (0\columnA, 0\rowA) { Input signal $f$ };
			\node [style={rect2}] (n5) at (0.6\columnA, 1\rowA) { Noise $\noise$ };
			\node [style={rect2}] (n6) at (0.6\columnA, -1\rowA) { True signal $\bar{f}$ };
			\node [style={rect18}, scale = 2.0] (n4) at (1.2\columnA, 0\rowA) { $B_{N, \delta_3}$ };
			\node [text width=0.22\columnA, text centered, minimum height=0.2\rowA, shape=rectangle, draw=black, fill=ChhaiC, scale = 2.0] (n3) at (1.9\columnA, 0\rowA) { $\Matrix{ K^{\nu} }$ };
			\node [text width=0.45\columnA, text centered, minimum height=0.2\rowA, shape=rectangle, draw=black, fill=ChhaiB, scale = 2.0] (n2) at (2.8\columnA, 0\rowA) { $ \SqBrack{ A_{N, \delta_3, M} + \theta }^{-1}$ };
			\node [style={rect15}, scale = 2.0] (n1) at (3.7\columnA, 0\rowA) { $K^{\nu}$ };
			\node [text width=0.3\columnA, text centered, minimum height=0.2\rowA, shape=rectangle, draw=black, fill=white, scale=2.0] (n8) at (4.4\columnA, 0\rowA) { Output };
			\node [text width=1.5\columnA, text centered, minimum height=0.2\rowA, shape=rectangle, draw=ChhaiB] (n9) at (2.4\columnA, -2.5\rowA) { \textbf{2.} Sampling measure $\nu$ depends on subgrid size $M$. $M$ is chosen so that $\bar{f}$ can be approximated $\epsilon_1$-closely by a function in $\rkhs_\nu$ };
			\node [text width=1.2\columnA, text centered, minimum height=0.2\rowA, shape=rectangle, draw=ChhaiB] (n11) at (0.6\columnA, -2.5\rowA) { \textbf{1.} Bandwidth $\delta_3$ chosen so that enough samples of noise $\noise_{\nu}$ \eqref{eqn:def:err:1} is aggregated to get a sum less than $\epsilon_2$ with high probability. };
			\node [text width=1.2\columnA, text centered, minimum height=0.2\rowA, shape=rectangle, draw=ChhaiB] (n12) at (4\columnA, -2.3\rowA) { \textbf{3.} Regularization parameter $\theta$ chosen so that the contribution of the averaged noise to the regression problem is suppressed. See \eqref{eqn:def:err:6}. };
			\node [style={rect2}] (n13) at (1.2\columnA, 2.5\rowA) { Markov kernel smoothing };
			\node [style={rect2}] (n14) at (1.9\columnA, 2.5\rowA) { RKHS kernel matrix };
			\node [style={rect2}] (n15) at (2.8\columnA, 2.5\rowA) { regularized RKHS regression };
			\node [style={rect2}] (n16) at (3.7\columnA, 2.5\rowA) { RKHS kernel smoothing };
			\draw[-to] (n7) to (n6);
			\draw[-to] (n7) to (n5);
			\draw[-to] (n5) to (n4);
			\draw[-to] (n6) to (n4);
			\draw[-to] (n4) to (n3);
			\draw[-to] (n3) to (n2);
			\draw[-to] (n2) to (n1);
			\draw[-to] (n1) to (n8);
			\draw[line width=1pt, rounded corners] (n9) -- (2.4\columnA, -2\rowA) -- (1.9\columnA, -2\rowA) -- (n3);
			\draw[line width=1pt, rounded corners] (n11) -- (0.6\columnA, -1.6\rowA) -- (1.2\columnA, -1.6\rowA) -- (n4);
			\draw[line width=1pt, rounded corners] (n12) -- (4\columnA, -1.5\rowA) -- (2.8\columnA, -1.5\rowA) -- (n2);
			\draw[-to] (n13) to (n4);
			\draw[-to] (n14) to (n3);
			\draw[-to] (n15) to (n2);
			\draw[-to] (n16) to (n1);
		\end{tikzpicture}
		\caption{Kernel based denoising as a linear filter. The net transformation effected by Algorithm \ref{algo:1} is linear, as dilineated in \eqref{eqn:scheme:5}. We depict this transformation as a sequence of matrices. The construction of the four matrices are defined in \eqref{eqn:scheme:5}. Suppose $\epsilon_1, \epsilon_2 > 0$ are two error bounds, and $\beta_X$ is the sampling measure created from a fixed grid of $N$ pixels. The diagram presents how the various parameters are chosen so that the error bounds are satisfied. The choice of some parameters are dependent on the choice of others. The numbering represents the order in which these parameters are chosen.}
		\label{fig:linear_filter}
	\end{figure}
	
	\begin{algo} \label{algo:3}
		Denoising in patches
		\begin{itemize}
			\item \textbf{Input.} A grayscale image of dimensions $l\times w$ and pixels $m\times n$.
			\item \textbf{Parameters.} 
			\begin{enumerate}
				\item All the parameters of Algorithm \ref{algo:1}.
				\item A zero-threshold $\epsilon_{zero} \approx 10^{-16}$.
			\end{enumerate}
			\item \textbf{Output.} A denoised image.
			\item \textbf{Steps.}
			\begin{enumerate}
				\item Compute the minimum radius $R>0$ such that the disk of radius $R$ centered at the origin of the plane $\real^2$, contains at least $\eta^2$ lattice point.
				\item Set $\delta = R^2 / \log \paran{ \epsilon_{zero} }$.
				\item Set $\hat{f}=0$.
				\item Repeat for $i=1,\ldots,m$
				\item \quad Compute the estimate $\hat{f}_i$ on the patch $\calU_i$ using Algorithm \ref{algo:1}.
				\item \quad Compute the function $\mathfrak{s}$ on $\calU_i$, using \eqref{eqn:aed54g}.
				\item \quad For each pixel location $x\in \calU_i$, increment the value of $\hat{f}(x)$ by $\hat{f}_i (x) \phi \paran{x; \calU_i, h_i} / \mathfrak{s}(x)$.
			\end{enumerate}
		\end{itemize}
	\end{algo}
	
	Algorithm \ref{algo:3} explains how the core denoising algorithm \ref{algo:1} can be applied separately on each patch, and then the results aggregated. This completes the description of our data-driven procedure. Note that there are many parameters in the algorithms which are not preset, and need to be chosen according to the task. These are the various bandwidth parameters $\delta_2, \delta_3$, the RKHS kernel $k$ and Markov kernel $p$. 
	
	We next examine the convergence properties of our technique.	
	
	\section{Convergence analysis} \label{sec:cnvrgnc}
	
	Algorithm \ref{algo:1} implements the least-squares regression problem posed in \eqref{eqn:scheme:6}. Equation \eqref{eqn:scheme:6} is the data-driven version of identity \eqref{eqn:scheme:2}, which equates the a kernel integral transform on the noise-free image, to a kernel integral transform of the noisy image. As described in Section \ref{sec:algo}, $\beta$ and $\beta_X$ is completely characterized by the pixel count $N$, $\nu$ is characterized by a sub-sampling parameter $M$. Thus the output $\hat{f}_{\beta, \nu, \theta}$ of Algorithm \ref{algo:1} can be written as  
	\begin{equation} \label{eqn:scheme:5}
		\begin{split}
			\hat{f}_{N, M, \theta} := \hat{f}_{\beta, \nu, \theta} &:= \Shobuj{ \Matrix{ K^{\nu} } }
			\akashi{ \SqBrack{ \Matrix{ K^{\nu} }^T \itranga{ \Matrix{ P^{\beta_X}_{\delta_3} }^T \Matrix{ P^{\beta_X}_{\delta_3} } } \Matrix{ K^{\nu} } + \theta }^{-1} }
			\Shobuj{ \Matrix{ K^{\nu} }^T } \itranga{ \Matrix{ P^{\beta_X}_{\delta_3} }^T
				\Matrix{ \tilde{P}_{\delta_3}^{\beta} } }
			f \\
			&= \Shobuj{ \Matrix{ K^{\nu} } } \akashi{ \Matrix{ A_{N, \delta_3, M} + \theta }^{-1} } \Shobuj{ \Matrix{ K^{\nu} } } \itranga{ \Matrix{ B_{N, \delta_3} } } f
		\end{split}
	\end{equation}
	where we have set
	\[ B_{N, \delta_3} := \Matrix{ P^{\beta_X}_{\delta_3} }^T \Matrix{ P^{\beta_X}_{\delta_3} } , \quad A_{N, \delta_3, M} := \Matrix{ K^{\nu} } B_{N, \delta_3} \Matrix{ K^{\nu} } . \]
	The identity in \eqref{eqn:scheme:6} leads to the following almost-sure convergence for \eqref{eqn:scheme:5} :
	\begin{equation} \label{eqn:scheme:7}
		\bar{f} \in \rkhs \imply \boxed{ \lim_{M\to \infty} \lim_{N\to \infty, \theta\to 0^+} \norm{ \hat{f}_{N, M, \theta} - \bar{f} }_{\sup} \stackrel{a.s.}{=} 0. }
	\end{equation}
	The almost surety stems from the fact that the pixels and noise samples are equidistributed with respect to $\mu$. 
	Recall that the function $\hat{f}_{N, M, \theta}$ is created by the kernel smoothing operation defined in \eqref{eqn:def:os}. Thus the function $\hat{f}_{N, M, \theta}$ is completely data-driven. Equation \eqref{eqn:scheme:7} establishes the convergence of our data-driven technique in the limit of infinitely many pixels sampling the underlying image. We next investigate the accuracy of the algorithm for finitely pixellated data.
	
	\paragraph{Parameter tuning} Equation \eqref{eqn:scheme:5} reveals that the overall transformation is a sequence of linear transformations
	The construction of each matrix comes along with the choice of four parameters - 
	\begin{enumerate} [(i)]
		\item a sampling measure $\beta$ of $\calX\times \calY$, created from the pixel information of the noisy image.
		\item a sub-sampling measure $\nu$ of $\beta_X$, created from a choice of a sub-sampling parameter.
		\item a bandwidth parameter $\delta_3$ that sets the  effective local radius of averaging.
		\item a regularization parameter $\theta>0$ for the linear regression problem.
	\end{enumerate}
	
	The choice of these parameters depend on the following three considerations -- 
	\begin{enumerate} [(i)]
		\item the profile of the noise $\noise$, more specifically, its central tendency.
		\item the smoothness of the noise-free image $\bar{f}$ in terms of the RKHS created by the kernel $k$.
		\item the granularity of the image, i.e., the number of pixels.
	\end{enumerate}
	
	We combine all these considerations in a different analysis of the convergence established in \eqref{eqn:scheme:7}. This time, we shall lose the assumption that $\bar{f}\in \rkhs$ and instead utilize the density of $\rkhs$ within $C(\calX)$. The analysis proceeds according to the following logical steps :
	
	\paragraph{Separating noise} Since we have replaced $\mu$ with $\beta$ and $\mu_X$ with $\nu$, the equality \eqref{eqn:scheme:1} is no longer satisfied. In that case the following difference becomes relevant :
	\begin{equation} \label{eqn:def:err:1}
		\noise_{\beta} := \paran{ \tilde{P}^{\beta} - \tilde{P}^{\mu} }f + \paran{ P^{\beta_X} - P^{\mu_X} } \bar{f} .
	\end{equation}
	As a result we have
	\begin{equation} \label{eqn:def:err:2}
		\tilde{P}^{\beta} f = P^{\beta_X} \bar{f} + \noise_{\beta}.
	\end{equation}
	The LHS of \eqref{eqn:def:err:2} is the RHS of regression problem being solve in \eqref{eqn:scheme:4} and \eqref{eqn:scheme:5}. Thus \eqref{eqn:def:err:2} expresses the RHS of the regression problem as an integral transform $P^{\beta_X} \bar{f}$ of the true mean $\bar{f}$, along with the noise effect $\noise_{\beta}$ defined in \eqref{eqn:def:err:1}. It also provides the following alternate definition of $\noise_{\beta}$ :
	\begin{equation} \label{eqn:def:err:7}
		\paran{\noise_{\beta}}_j = \frac{1}{N} \sum_{n=1}^{N} p(x_j, x_n) y_n, \quad 1\leq j \leq N.
	\end{equation}
	Equations \eqref{eqn:def:err:1}, \eqref{eqn:def:err:2} and \eqref{eqn:def:err:7} are all equivalent definitions of the noise-residual vector.

	\paragraph{Approximation 1.} Henceforth we shall assume a fixed number of pixels $N$ distributed evenly over the domain $\calX$. This leads to an RKHS subspace
	\[ \rkhs_N := \spn \SetDef{ k(x_n, \cdot) }{ 1\leq n \leq N } . \]
	Fix an error level $\epsilon_1$. Let $B(\calX)$ be the collection of bounded functions on $\calX$. Consider the set
	\[ \rkhs_N(\epsilon_1) := \SetDef{ \hat{f} \in B(\calX) }{ \exists h\in \rkhs_N \mbox{ s.t. } \norm{ \hat{f} - h }_{B(\calX)} < \epsilon_1 } . \]
	Thus $\rkhs_N(\epsilon_1)$ contains those bounded functions which can be uniformly approximated within an error limit of $\epsilon_1$ by an RKHS function in $\rkhs_N$. Also note that
	\begin{equation} \label{eqn:approx:1}
		\clos \cup_{N\in\num} \rkhs_N(\epsilon_1) = B(\calX), \quad \forall \epsilon_1>0.
	\end{equation}
	We shall assume henceforth that the true image $\bar{f}$ lies in $\rkhs_N(\epsilon_1)$. Thus there is a vector $\hat{a}\in L^2(\nu)$ such that $\norm{ \bar{f} - K^\nu \hat{a} } < \epsilon_1$. Allowing a limit of $\epsilon_1$ we shall assume without loss of generality that $\bar{f} = K^\nu \bar{a}$.
	
	\begin{table}[!t]
		\caption{Parameters which remain constant in the experiments. }
		\begin{tabularx}{\linewidth}{|L|L|L|L|} \hline
			Variable & $\eta_3$ & $(l, w)$ & $\theta$ \\ \hline
			Interpretation & Parameter for selecting the bandwidth $\delta_3$ of the Markov kernel, based on Algorithm \ref{algo:2} & Dimensions of the image & Ridge regression parameter, as described in \eqref{eqn:def:ridge} \\ \hline
			Value & $4$ & $(1,1)$ & $0.01 \times \norm{K^\nu}$ \\ \hline
		\end{tabularx}
		\label{tab:param2}
	\end{table}
	
	\paragraph{Approximation 2} Equations \eqref{eqn:scheme:5}, \eqref{eqn:def:err:1} and \eqref{eqn:def:err:2} together imply
	\begin{equation} \label{eqn:def:err:3}
		\begin{split}
			\hat{f}_{N, M, \theta} &= \Shobuj{ K^{\nu} } \akashi{ \SqBrack{ \Matrix{ K^{\nu} }^T \itranga{ \Matrix{ P^{\beta_X}_{\delta_3} }^T \Matrix{ P^{\beta_X}_{\delta_3} } } \Matrix{ K^{\nu} } + \theta }^{-1} } \Shobuj{ \Matrix{ K^{\nu} }^T } \itranga{ \Matrix{ P^{\beta_X}_{\delta_3} }^T } \SqBrack{ P^{\beta_X} \bar{f} + \noise_{\beta} } \\
			&= \Shobuj{ K^{\nu} } \akashi{ \SqBrack{ A_{N, \delta_3, M} + \theta }^{-1} } \Shobuj{ \Matrix{ K^{\nu} } } \SqBrack{ B_{N, \delta_3}\bar{f} + \Matrix{ P^{\beta_X}_{\delta_3} }^T \noise_{\beta} } .
		\end{split}
	\end{equation}
	Incorporating the assumption $\bar{f} = K^\nu \bar{a}$ gives
	\begin{equation} \label{eqn:def:err:4}
		\hat{f}_{N, M, \theta} = \Shobuj{ K^{\nu} } \akashi{ \SqBrack{ A_{N, \delta_3, M} + \theta }^{-1} } \akashi{ \SqBrack{ A_{N, \delta_3, M} } } \bar{a} + \Shobuj{ K^{\nu} } \akashi{ \SqBrack{ A_{N, \delta_3, M} + \theta }^{-1} } \Shobuj{ \Matrix{ K^{\nu} } } \Matrix{ P^{\beta_X}_{\delta_3} }^T \noise_{\beta} .
	\end{equation}
	Thus $\hat{f}_{N, M, \theta}$ is the sum of two separate and mutually independent terms, one which depends on the true image and one which only depends on the noise profile. We shall examine these terms separately.
	
	\paragraph{Approximation 3} Note that the limit in \eqref{eqn:scheme:7} holds irrespective of the nature of $\mu$. Suppose $\mu$ is the inclusion of $\mu_X$ in $\calX \times \calY$, i.e., for each $x\in \calX$, $\nu_{\calY|x} = \delta_0$. In that case according to \eqref{eqn:def:err:1}, $\noise_{\nu} \equiv 0$. Thus \eqref{eqn:scheme:7} and \eqref{eqn:def:err:4} combine to give
	\[\lim_{N\to \infty, \theta\to 0^+} \norm{ \SqBrack{ A_{N, \delta_3, M} + \theta }^{-1} \SqBrack{ A_{N, \delta_3, M} } \bar{a} - \bar{a} }_{L^2(\nu)} = 0. \]
	Since we keep the sub-sampling parameter $M$ fixed and thus $\nu$ fixed, $K^\nu$ remains a fixed and bounded operator. Thus :
	\begin{equation} \label{eqn:def:err:5}
		\lim_{N\to \infty, \theta\to 0^+} \norm{ K^{\nu} \SqBrack{ A_{N, \delta_3, M} + \theta }^{-1} \SqBrack{ A_{N, \delta_3, M} } \bar{a} - K^{\nu} \bar{a} }_{C(\calX)} = 0. 
	\end{equation}
	One important property of the convergence in \eqref{eqn:def:err:5} is that $N, \theta$ can be simultaneously taken to their limiting values. This joint convergence is one of the distinguishing feature of our technique.
	
	\paragraph{Approximation 4} The self adjoint operator $B_{\beta_{X}, \delta_3} = \paran{P^{\beta_{X}}_{\delta_3}}^T P^{\beta_{X}}_{\delta_3}$ is also a kernel integral operator with kernel
	\[ b^{\beta_{X}}_{\delta_3} : \calX \times \calX \to \real, \quad (z, x') \mapsto \int_{\calX} p(z, x') p(x, x') d\beta_{X}(x) \]
	Then according to \eqref{eqn:def:err:2}, we have
	\[\begin{split}
		\SqBrack{ \paran{ P^{\beta_X}_{\delta_3} }^T \noise_\beta }(z) &= \int_{\calX} b^{\beta_{X}}_{\delta_3}(z, x') \SqBrack{ \int_{\calY} f(x', y) d \beta_{\calY|x'}(y) - \bar{f}(x') } d\beta_X(x') \\
		&= \int_{\calX} b^{\beta_{X}}_{\delta_3}(z, x') \SqBrack{ \int_{\calY} f(x', y) d \beta_{\calY|x'}(y) - \int_{\calY} f(x', y) d \mu_{\calY|x'}(y) } d\beta_X(x')\\
		&= \int_{\calX} b^{\beta_{X}}_{\delta_3}(z, x') \int_{\calY} \SqBrack{ f(x', y)- \bar{f}(x') } d \beta_{\calY|x'}(y) d\beta_X(x') .
	\end{split}\]
	We now assume the additive noise model of \eqref{eqn:def:img_noise}. Then
	\[ \SqBrack{ \paran{ P^{\beta_X}_{\delta_3} }^T \noise_\beta }(z) = \frac{1}{N} \sum_{n=1}^{N} b^{\beta_{X}}_{\delta_3}(z, x_n) y_n \]
	%
	where $\vec{y} = \paran{ y_1, \ldots, y_N}$ is the vector of noise values added to the pixels. By the assumption of zero-mean noise, and by the central limit theorem we have
	\[\begin{tikzcd}
		X(N,z, \delta_3) := N^{1/2} \SqBrack{ \paran{ P^{\beta_X}_{\delta_3} }^T \noise_\beta }(z) \arrow{rr}{ a.s. }[swap]{N\to\infty} && \calN \paran{ 0, \text{var}(y) } .
	\end{tikzcd}\]
	This almost sure convergence holds uniformly for all locations $z\in \calX$. We have declared the quantity to be a random variable $X(N,z, \delta_3)$. Let us fix an error bound $\epsilon_4$. Thus by the Chebyshev inequality we have
	\[ \liminf_{N\to\infty} \text{Prob} \paran{ \abs{ X(N,z, \delta_3) } < \epsilon_4^{-1/2} } > 1 - \epsilon_4.\]
	This allows us to bound the second term in \eqref{eqn:def:err:4} as
	\begin{equation} \label{eqn:def:err:6}
		\begin{tikzcd}
			\limsup_{N\to\infty} \text{Prob} \paran{ \abs{ K^{\nu} \SqBrack{ A_{N, \delta_3, M} + \theta }^{-1} \Matrix{ K^{\nu} } \Matrix{ P^{\beta_X}_{\delta_3} }^T \noise_{\beta} } < \frac{1}{ \theta N^{1/2} \epsilon_4^{1/2} } } > 1-\epsilon_4 .
		\end{tikzcd}
	\end{equation}
	Equations \eqref{eqn:def:err:4}, \eqref{eqn:def:err:5} and \eqref{eqn:def:err:6} imply :
	
	\begin{theorem} \label{thm:2}
		Suppose Assumptions \ref{A:1} and \ref{A:4} hold, and $\bar{f} : \calX \to \real$ be a bounded function representing a noise-free grayscale image. Let $a_{N, M, \theta}$ denote the output of Algorithm~\ref{algo:1} is applied to the data $\paran{ x_n, y_n }_{n=1}^{N}$. Fix error bounds $\epsilon_1, \epsilon_2, \epsilon_3 , \epsilon_4>0$. 
		\begin{enumerate} [(i)]
			\item For $M$ large enough, $\bar{f}$ may be approximated uniformly within an error of $\epsilon_1$ as a sum of kernel sections 
			\[ \norm{ \bar{f} - \sum_{m=1}^{M} \bar{a}_m k \paran{ \tilde{x}_m, \cdot } }_{ \sup \calX } < \epsilon_1 . \]
			Here $\braces{ \tilde{x} }_{m=1}^{M}$ is a grid of uniformly distributed points on $\calX$.
			\item Given such a choice of $M$, there exists $\theta_0>0$ and $N'_0\in\num$ such that for every $N>N'_0$ and $\theta \in (0, \theta_0)$, the first term in \eqref{eqn:def:err:4} is less than $\epsilon_2$.
			\item Fix such a $\theta \in (0, \theta_0)$. Then there is $N_0>N_0 \in\num$ such that for every $N>N_0$, with probability at least $1-\epsilon_4$, the second term in \eqref{eqn:def:err:4} is less than $\frac{1}{ \theta N^{1/2} \epsilon_4^{1/2} }$.
		\end{enumerate}
	\end{theorem}
	
	In other words, 
	
	\begin{corollary} \label{corr:3}
		Let $\bar{f} : \calX \to \real$ be a bounded function representing a noise-free grayscale image on a compact domain $\calX$. Given $\epsilon_1, \epsilon_2, \epsilon_3$, for $N,M$ large enough and $\theta>0$ small enough, with probability at least $1-\epsilon_4$ the estimate $\hat{f}_{N, M, \theta}$ from \eqref{eqn:scheme:5} satisfies :
		\begin{equation} \label{eqn:scheme:8}
			\norm{ \hat{f}_{N, M, \theta} - \bar{f} }_{\sup \calX} < \epsilon_1 + \epsilon_2 + \epsilon_3 .
		\end{equation}
	\end{corollary}
	
	The strategy presented in Figure \ref{fig:linear_filter} and in Theorem \ref{thm:2} for choosing the parameters applies to general kernel based denoising tasks \cite[e.g.]{Das2023conditional, Das2024drift}. We next describe some numerical experiments to test our technique on.
	
	\section{Examples} \label{sec:examples}  
	
	\begin{figure}[!t]
		\centering
		\includegraphics[width=.45\linewidth]{\figs 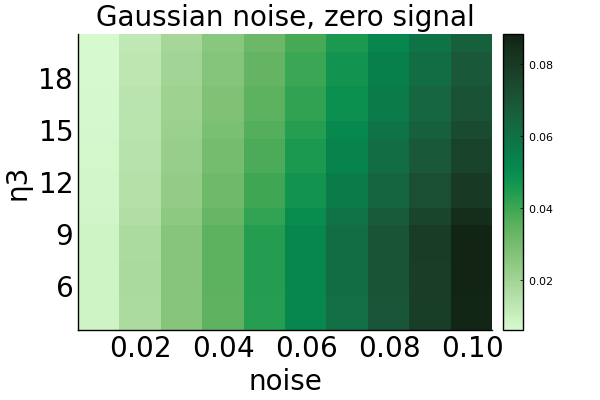}
		\includegraphics[width=.45\linewidth]{\figs 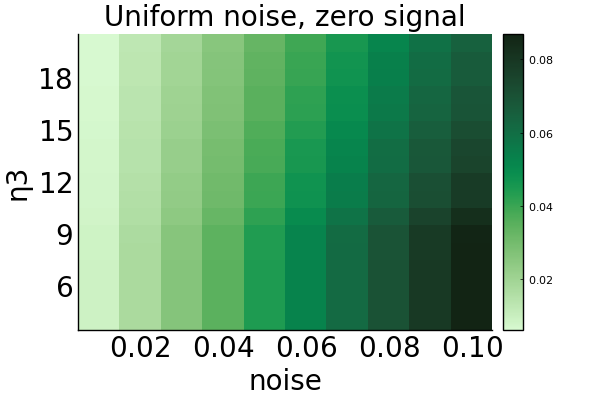}
		\caption{Results of experiment 1. The performance of the denoiser has been checked on an zero-signal noise, i.e., an $m\times n$ image in which $\bar{f} \equiv 0$. The graph shows the L2 error for various choices of $\eta_3$. See Table \ref{tab:param2} for a list of the various parameters of the algorithm which are kept constant. }
		\label{fig:zero_noise}
	\end{figure}
	
	Algorithm \ref{algo:1} requires a choice of a markov kernel $p$ , and an RKHS kernel $k$. The convergence results of Lemma \ref{lem:scheme:6} and equation \eqref{eqn:scheme:6} are independent of these choices. We now describe our choice for these kernels for the numerical experiments.
	
	\paragraph{Markov kernel} As described in \eqref{eqn:def:GaussMrkv}, our markov kernel is obtained by the markov normalization of a gaussian kernel. The bandwidth of this Gaussian kernel will be denoted as $\delta_3$. We explain later how the choice of $\delta_3$ is made.
	
	\paragraph{RKHS kernel} Our choice of kernel in all the experiments is the \emph{diffusion} kernel  \cite[e.g.][]{MarshallCoifman2019, WormellReich2021}. Among its various constructions, we choose the following :
	\begin{equation} \label{eqn:def:kdiff}
		\begin{split}
			& k_{ \text{diff}, \delta_2}^\mu (x,y) = \frac{ k_{\text{Gauss}, \delta_2}(x,y) }{ \deg_l(x) \deg_r(y) } ,\\ 
			& \deg_r(x) := \int_X k_{\text{Gauss}, \delta_2}(x,y) d\mu(y) , \quad \deg_l(x) := \int_X k_{\text{Gauss}, \delta_2}(x,y) \frac{1}{ \deg_r(x) } d\mu(y) .
		\end{split}
	\end{equation}
	Diffusion kernels have been shown to be good approximants of the local geometry in various different situations \citep[e.g.][]{CoifmanLafon2006, HeinEtAl2005, VaughnBerryAntil2019, DGGS2024sec}, and are a natural choice for non-parametric learning. It has the added advantage of being symmetrizable :
	\begin{equation} \label{eqn:symm}
		\rho(x) k_{ \text{diff}, \delta_2}^\mu (x,y) \rho(y)^{-1} = \tilde{k}_{ \text{diff}, \delta_2}^\mu (x,y) = \frac{k_{\text{Gauss}, \delta_2}(x,y)}{ \SqBrack{ \deg_r(x) \deg_r(y) \deg_l(x) \deg_l(y) }^{1/2} }, 
	\end{equation}
	where
	\[ \rho(z) = \deg_l(z)^{1/2} / \deg_r(z)^{1/2} . \]
	The kernel $\tilde{k}_{ \text{diff}, \delta_2}^\mu$ from \eqref{eqn:symm} is clearly symmetric. Since it is built from the symmetric positive definite (s.p.d.) kernel $k_{ \text{Gauss}, \delta_2}$, $\tilde{k}_{ \text{diff}, \delta_2}^\mu$ is s.p.d. too and thus generates an RKHS of its own. Moreover, the kernel $k_{ \text{diff}, \delta_2}^\mu$  can  be symmetrized by a degree function $\rho$, which is both bounded and bounded above $0$. Such a kernel will be called \emph{RKHS-like}. Let $M_\rho$ be the multiplication operator with $\rho$. Then
	\[ \ran K^\mu_{ \text{diff}, \delta_2} = \ran M_\rho \circ \tilde{K}^\mu_{ \text{diff}, \delta_2} . \]
	Again, because of the properties of $\rho$, both $M_\rho$ and its inverse are bounded operators. Thus there is a bijection between the RKHS generated by $\tilde{k}_{ \text{diff}, \delta_2}^\mu$, and the range of the integral operator $K_{ \text{diff}, \delta_2}^\mu$. 
	
	The choice of the RKHS kernel thus depends on the choice of the bandwidth $\delta_2$. This quantity is independent and separate from the Markov kernel bandwidth $\delta_3$. We next describe a common procedure for deciding both the bandwidths $\delta_2$ and $\delta_3$.

	\paragraph{Bandwidth selection} Bandwidth selection is a crucial part of kernel methods, but there is no complete understanding about what would constitute an optimal bandwidth. Bandwidth is often based on a criterion for exclusion. A Gaussian kernel, along with all its normalized forms undergoes an exponential decay away from the diagonal. Thus given any zero threshold $\epsilon_{zero}$, there is a radius $R$ such that the kernel value $k(x,y)$ is less than $\epsilon_{zero}$ if $\dist(x,y)>R$. One way of choosing the bandwidth is based on controlling this radius $R$. The choice of the effective radius $R$ also determines how many data-points get aggregated by kernel calculations based at a point. We select the bandwidth based on a predetermined integer $\eta$. This number $\eta$ represents the number of neighboring pixels in either directions that we wish to incorporate in the kernel computations. Given a choice $\eta$, the bandwidth $\delta$ is computed as follows : 
	
	\begin{algo} \label{algo:2}
		Bandwidth selection
		\begin{itemize}
			\item \textbf{Input.} 
			\begin{enumerate}
				\item Length and width $(l,w)$ of the image.
				\item Number of pixels $(m,n)$ along the length and width respectively.
				\item An integer $\eta$.
			\end{enumerate}
			\item \textbf{Parameters.} A threshold $\epsilon_{zero}$ for zero , typically $10^{-14}$.
			\item \textbf{Output.} A bandwidth $\delta>0$.
			\item \textbf{Steps.}
			\begin{enumerate}
				\item Compute the minimum radius $R>0$ such that the disk of radius $R$ centered at the origin of the plane $\real^2$, contains at least $\eta^2$ lattice point.
				\item Set $\delta = R^2 / \log \paran{ \epsilon_{zero} }$.
			\end{enumerate}
		\end{itemize}
	\end{algo}
	
	The auto-tuning procedures displayed in Figure \ref{fig:outline:2} use Algorithm \ref{algo:2}, for certain choices of $\eta_2, \eta_3$ for $\delta_2, \delta_3$ respectively. We next describe our experiments.
	
	\begin{figure}[!t]
		\centering
		\includegraphics[width=.31\linewidth]{\figs 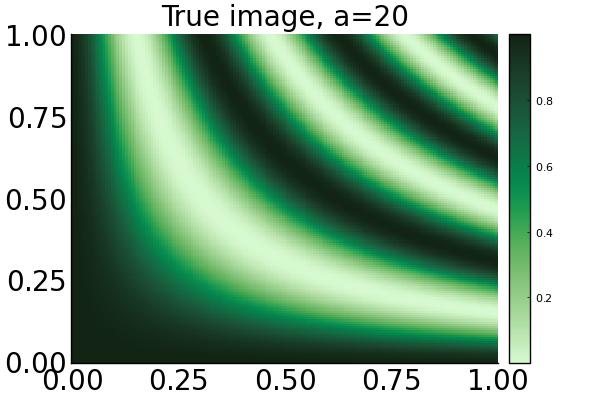}
		\includegraphics[width=.31\linewidth]{\figs 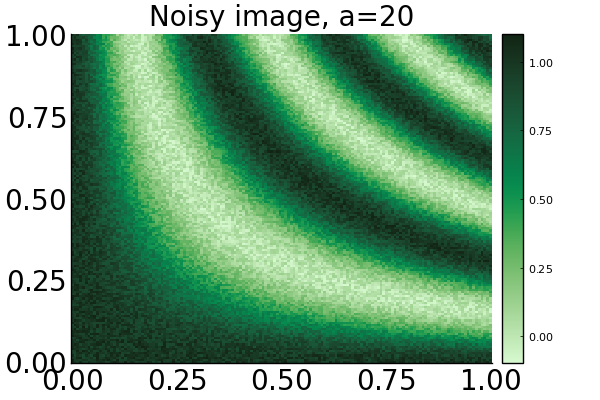}
		\includegraphics[width=.31\linewidth]{\figs 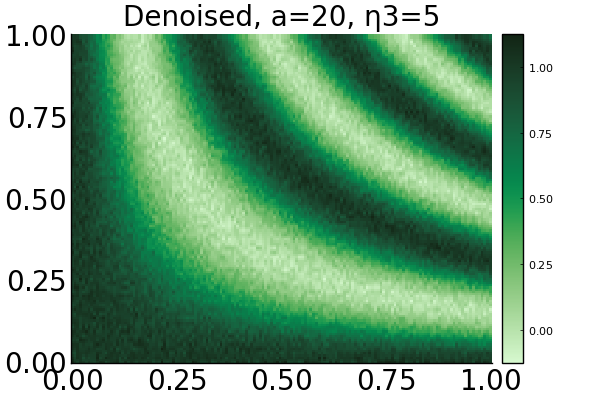}
		
		\includegraphics[width=.31\linewidth]{\figs 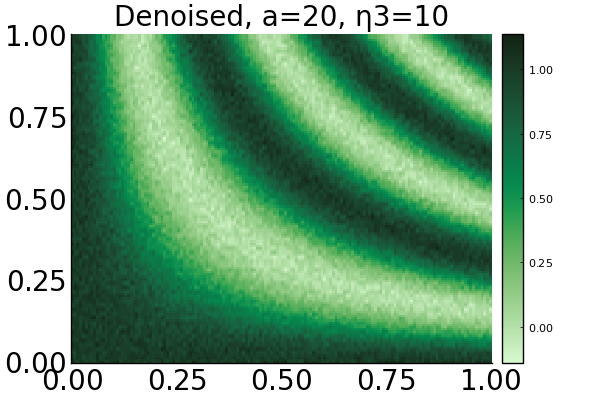}
		\includegraphics[width=.31\linewidth]{\figs 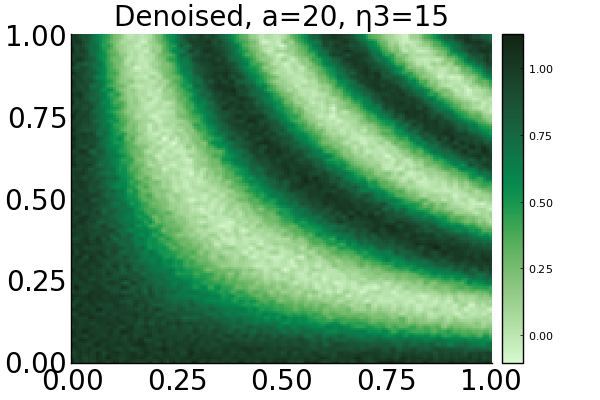}
		\includegraphics[width=.31\linewidth]{\figs 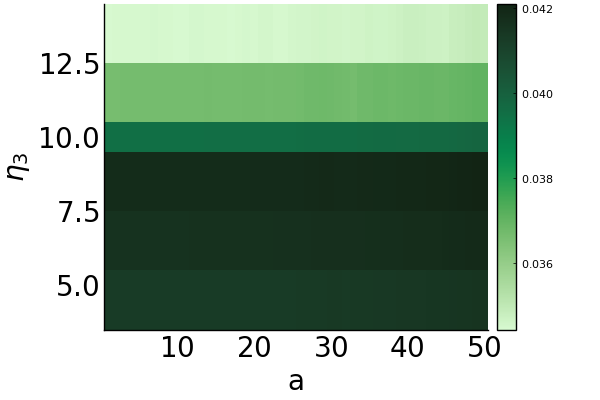}
		\caption{Results of experiment 2. The first five panels show the image described by the function \eqref{eqn:expt:1} at $\alpha=20$, and a Gaussian random noise of standard deviation $0.1$. The parameter $\alpha$ controls the smoothness of the image. A hither value of $\alpha$ means a less smooth image, which is also harder to denoise. The panels show the result of denoising at various values of the algorithmic parameter $\eta_3$. A higher value of $\eta_3$ implies a larger local area over which the pixels are averaged out. This leads to a better denoising. The last panel shows the performance of denoising as a function of $\alpha$ ad $\eta_3$.}
		\label{fig:eta_v_a}
	\end{figure}
	
	\paragraph{Experiments} In all the experiments we use a parameter $\sigma$ as an index of the variance of the noise model. One of the models we use is a uniform noise model
	\begin{equation} \label{eqn:noise_uni}
		\mu_Y = \mbox{ uniform probability measure on } (-\sigma, \sigma).
	\end{equation}
	In some other experiments we use a normally distributed noise : 
	\begin{equation} \label{eqn:noise_normal}
		\mu_Y = \sim \calN( 0, \sigma ).
	\end{equation}
	The overall goal of our experiments is to test the efficacy of our methods in image denoising. For that we test Algorithm \ref{algo:1} on a images of varying degree of smoothness and pixel count. 
	\begin{enumerate}
		\item Since our denoising technique is overall a linear procedure, we begin by testing the technique on a zero-signal image. In terms of Assumptions \ref{A:1} and \ref{A:2}, it means that the grayscale function $\bar{f}$ is identically zero. See Figure \ref{fig:zero_noise} for the results.
		\item Next we consider a $250 \times 250$ image, whose noise-free grayscale value is given by
		\begin{equation} \label{eqn:expt:1}
			\bar{f}(x_1, x_2) = 0.5\SqBrack{ 1+cos(\alpha x_1)*cos(\alpha x_2) }, \quad \forall (x_1, x_2) \in [0,l] \times [0,w] .
		\end{equation}
		The parameter $\alpha$ is an experimental parameter controlling the smoothness of the image. See Figure \ref{fig:eta_v_a} for results.
		\item A real-world image in Figure \ref{fig:shundorbon2:a} in which there is a continuous background, continuous foreground, and piecewise smooth intermediary object.
		\item A real-world image in Figure \ref{fig:howrah_bridge:a} with a geometric network like structure amidst a smooth backdrop.
		\item A real-world image in Figure \ref{fig:jhor:a} with objects with fractal boundaries amidst a smooth backdrop.
	\end{enumerate}
	
	The only algorithmic parameter that we adjust is $\eta_3$. This is an integer parameter controlling the effecting number of pixels whose noise values get averaged locally under the Markov kernel. Table \ref{tab:param2} lists those algorithmic parameters which are kept constant in the experiments. 	
	
	\section{Conclusions} \label{sec:conclus}
	
	In this paper, we developed a data-driven statistical image denoising technique using the RKHS framework. More specifically, we constructed an unsupervised learning method to denoise images within an RKHS. We described the denoised image as the solution to a least squares regression problem, where the true image is the conditional expectation of a noisy image. Our entire procedure was presented through three algorithms, illustrating how regression can be performed on image data, how denoising can be achieved using small patches (i.e., subcovers), and how the bandwidth parameter can be selected. A key feature of our method is that it accounts for different feature levels of the image during denoising. Alongside numerical experiments, we provided a convergence analysis that demonstrates both the theoretical and numerical efficiency of our approach.\\
	
	In summary we have shown the following: 
	\begin{enumerate}
		\item Boundary effects : An RKHS based learning technique that uses a smooth kernel reconstructs a target function as a smooth function. This results in errors at the boundary of objects within the picture. This is particularly visible in the fractal and non-smooth boundaries of Figures \eqref{fig:shundorbon2}, \eqref{fig:howrah_bridge}, and \eqref{fig:jhor}. An interesting challenge would be a merger of kernel techniques with the methods of Total variation minimization (TVM) \cite{rudin1992nonlinear} and anisotropic diffusion (AD) \cite[e.g.]{perona1990scale, BaiFeng2018image}. Both techniques focus on image denoising while preserving the image structures, such as edges. TVM performs denoising by minimizing the total variation of the image, while AD denoises the image based on gradient information. Various variations of TVM has been proposed, such as ones aimed at Gaussian noise \cite{louchet2014total}, and Poisson noise \cite{abergel2015total}, and more general nosie models \cite{chambolle2004algo}.
		
		\item Smoothing effect : Smoothing effect is an unavoidable effect   of most image denoising techniques. This is specially true in our method which averages pixel values across edges. This effect is most visible as a loss of contrast of the entire image. Although local averaging reduces sharp differences in tonal variations, however our use of an RKHS helps retain the finer details. This effect can be seen from the experiments outlined in Figures \eqref{fig:shundorbon2}, \eqref{fig:howrah_bridge}, and \eqref{fig:jhor}. The local averaging with a Gaussian kernel is also known as \emph{Gaussian smoothing} \cite{lindenbaum1994gabor, gabor1965info}. Smoothing suppresses noise, but it also changes the intensity variation of the underlying image. This suppresses, or even removes, detailed features of the original image. In this paper, we study wavelet-based denoising as a possible alternative to Gaussian smoothing. Wavelet-based denoising has the advantage over low-pass filtering that relevant detail information is retained, while small details, due to noise, are discarded.
		\item Computational complexity : Overall our method has two distinct phases : a training phase and an execution phase. For a fixed image resolution, the training phase only needs to be run once. The objects created in the training phase are independent of the image. The actual image is input only at the execution phase. 
		
		As shown in Figure \ref{fig:linear_filter}, the training phase involves the creation of kernel matrices \(G_{\delta}\), \(P^{\beta_{X}}_{\delta_3}\), \(K^{v}\), and a linear regression solver. The purpose of this solver is to compute the inverse of \([A_{N, \delta_3, M}+\theta]\). This matrix is sparse of very high dimensions ($mn \times mn$). We implement this solver using a randomized SVD. Note that, for the fixed patch size and sub-sampling parameters, the kernel matrix construction is a one-time operation and does not need to be repeated.
		
		At the end of the training phase, we solved the  to estimate the regression coefficients. \\
		The execution phase begins with solving a regularized linear least-square problem to learn the coefficients of the denoised image in terms of the kernel function.  An application of the kernel integral operator to the coefficients yields the denoised image. In an SVD based approach, this entire phase involves a series of matrix multiplications to the image vector. 
		
		For a fixed patch size \(m_0 \times n_0\), the kernel matrices require pairwise distance comparisons between $m0 n0$ points, leading to a total of $\bigO{ m_0^2 n_0^2 }$ operations. The SVD computation requires about $\bigO{ m_0^2 n_0^2 }$. The execution phase only involves matrix-vector multiplications, leading to $\bigO{ m_0 n_0}$ operations.
		
		\item Convergence guarantee : Our analysis does not start from the assumption of a grayscale image of a particular size. Rather we start from Assumptions \ref{A:1}, \ref{A:3} and \ref{A:2} to enable a notion of a noise-free image, and also convergence to it. The convergence is established in \eqref{eqn:scheme:7}, and reinterpreted in Theorem \ref{thm:2} in a probabilistic language. Equation \eqref{eqn:scheme:7} states in a precise way how as the number of noisy pixels \(N\) and subsamples \(M\) increases and the regularization parameter \(\theta\) tends to zero, the estimator \(\hat{f}_{N, M, \theta}\) converges almost surely and uniformly to the true image \(\bar{f}\). Overall this means that if we keep increasing the pixel count, the estimation $\hat{f}$ will converge to the true noise- free image.
		
		\item Scalability : The localized nature of kernel methods make them easily scalable for images of high pixel count. We have shown in Algorithm \ref{algo:3} how the denoising can be done in patches, and then aggregated. The size of the patches is chosen based on constraints of the computer. This scalability is not offered by techniques such as those which are $L^1$ or TV-minimization, whose objective functions are global in nature.
	\end{enumerate}
	
	Our strategy for denoising can be further developed in a number of interesting directions :
	\begin{enumerate}
		\item Choice of kernel : Identifying an optimal kernel has been an open research question for a long time \cite{narayan2021optimal, baraniuk1993signal, crammer2002kernel}. We chose the diffusion kernel as a trade-off between smoothness, edge preservation, and computational efficiency. We used fixed and uniform bandwidth values values \(\delta_2\) and \(\delta_3\). This also assigns equal sampling importance to all the samples. In future work, we aim to explore other kernels, such as variable bandwidth kernels, where the bandwidth can vary depending on the local properties, for example, the density or curvature of the data \cite{fan1995data, berry2016variable}. 
		\item Weighted averaging techniques - The strategy of \emph{kernel tapering} introduced by Blackman and Tukey \cite{babadi2014review, blackman1960measurement} may have interesting applications to our technique. This strategy makes the kernel matrix sparse while preserving positive semi-definiteness of the kernel matrix \cite{shen2014kernel}. Moreover, tapering helps control spectral leakage, which is significant in preserving sharp features such as edges during the denoising process \cite{harris2005use}. We want to incorporate this concept to extend our RKHS-based image denoising method.     
		\item Multi-level denoising : The linearity of our procedure makes it amenable to multi-level denoising \cite[e.g.]{chan2010multilevel, chan2008fast}. The idea is to progressively create approximate estimates of the true function, and re-apply the procedure to the residual for the next level of denoising. This requires a modification of the residual using a suitable chosen \emph{residual correction functional}. The design of such a functional remains a challenge.
	\end{enumerate}
	
	\bibliographystyle{\Path unsrt_inline_url}
	\bibliography{\Path References.bib,Ref.bib}
\end{document}